\documentclass[12pt,eqsecnum]{iopart}
\usepackage{geometry}                
\usepackage{bm}
\usepackage{amssymb}
\usepackage{mathrsfs} 
\usepackage{graphicx}
\usepackage{setstack}
\def\TODAY{22 December 2011; 22 February 2012}
\begin{document}
\title{Generic thin-shell gravastars}
\author{Prado Martin-Moruno$^1$, Nadiezhda Montelongo Garcia$^{2,3}$, Francisco S.~N.~Lobo$^3$, {\rm and} Matt Visser$^1$}
\address{$^1$ 
School of Mathematics, Statistics, and Operations Research,\\
Victoria University of Wellington, PO Box 600, Wellington 6140, New Zealand}
\address{$^2$
Departamento de F\'\i{}sica, Centro de Investigaci\'on  y Estudios avanzados del I.P.N.,
A.P. 14-700,07000 M\'exico, DF, M\'exico}
\address{$^3$
Centro de Astronomia
e Astrof\'{\i}sica da Universidade de Lisboa, Campo Grande, \\
Edif\'{i}cio C8 1749-016 Lisboa, Portugal}
\eads{
\mailto{prado@msor.vuw.ac.nz}, 
\mailto{nmontelongo@fis.cinvestav.mx},
\mailto{flobo@cii.fc.ul.pt},
\mailto{matt.visser@msor.vuw.ac.nz}
}
\begin{abstract}
We construct generic spherically symmetric thin-shell gravastars by using 
the cut-and-paste procedure. We take considerable effort to make the analysis as 
general and unified as practicable; investigating both the internal physics of the transition layer and its interaction with ``external forces'' arising due to interactions between the transition layer and the bulk spacetime.  Furthermore, we discuss both the dynamic and static situations. In particular, we consider ``bounded excursion'' dynamical configurations, and probe the stability of static configurations. For gravastars there is always a particularly compelling configuration in which the surface energy density is zero, while surface tension is nonzero. 


\bigskip
\noindent
\TODAY;  \LaTeX-ed \today.

\bigskip
\noindent
Published: JCAP {\bf03} (2012) 034.   doi:10.1088/1475-7516/2012/03/034

\bigskip
\noindent
ArXiv eprint: 1112.5253

\end{abstract}
\maketitle
\def\d{{\mathrm{d}}}
\def\P{{\cal{P}}}
\def\O{{\cal{O}}}
\def\g{{\bm{g}}}
\def\eqref#1{{(\ref{#1})}}
\hrule
\markboth{Generic thin-shell gravastars}{}
\tableofcontents
\bigskip
\hrule
\markboth{Generic thin-shell gravastars}{}
\newpage
\section{Introduction}

Gravastars (gravitational-vacuum stars) are hypothetical objects mooted as alternatives to standard Schwarzschild black holes~\cite{gr-qc/0012094, gr-qc/0109035}.  Typically the interior is some simple nonsingular spacetime geometry such as de~Sitter space, the exterior is some close approximation to the Schwarzschild geometry, and there is some complicated transition layer near the location where the event horizon would otherwise have been expected to form. Thus, in the traditional gravastar picture, the transition layer replaces both the de Sitter and the Schwarzschild horizons, and consequently the gravastar model has no singularity at the origin and no event horizon, as its surface is located at a radius slightly greater than the Schwarzschild radius. In this model, the quantum vacuum undergoes a
phase transition at or near the location where the event horizon is expected to form. Considerable attention has been devoted to these objects, and to closely related ``dark stars'', ``quasi black holes'', ``monsters'', ``black stars'', and the like~\cite{darkstar, quasi-black-hole, monster, black-star}. Related models, analyzed in a different context,
have also been considered by Dymnikova \cite{Dymnikova}. Some models use a continuous distribution of stress-energy, which on rather general grounds must be anisotropic in the transition layer~\cite{anisotropic, anisotropic2}. Other models idealize the transition layer to being a thin shell~\cite{gr-qc/0310107} and apply a version of the Sen--Lanczos--Israel junction condition formalism~\cite{junction_formalism}. In fact, the latter approaches have been extensively analysed in the literature, and applied to a wide variety of scenarios \cite{gravastar2,thinshell2}.  Several  criteria related to potential observability have been explored \cite{observations}.

The key point of the present paper is to develop an extremely general and robust framework that can quickly be adapted to wide classes of generic thin-shell gravastars. We shall consider standard general relativity, with gravastars that are spherically symmetric, with the transition layer confined to a thin shell. The bulk spacetimes (interior and exterior) on either side of the transition layer will be spherically symmetric and static but otherwise arbitrary.  (So the formalism is simultaneously capable of dealing with gravastars embedded in Schwarzschild, Reissner--Nordstr\"om, Kottler, or de~Sitter spacetimes, or even ``stringy'' black hole spacetimes.  Similarly the gravastar interior will be kept as general as possible for as long as possible.)  The thin shell (transition layer) will be permitted to move freely in the bulk spacetimes, permitting a fully dynamic analysis. This will then allow us to perform a general stability analysis, where gravastar stability is related to the properties of the  matter residing in the thin-shell transition layer. 

Many of the purely technical aspects of the analysis are very similar in spirit to that encountered in a companion paper analyzing thin-shell traversable wormholes~\cite{GLV} --- mathematically there are a few strategic sign flips --- but physically the current framework is significantly  different. Consequently we shall very rapidly find our analysis diverging from the traversable wormhole case. Further afield, we expect that the mathematical formalism developed herein will also prove useful when considering spacetime ``voids'' (manifolds with boundary)~\cite{voids}. 

This paper is organized in the following manner: In Section \ref{secII} we outline in great 
detail the general formalism of generic dynamic spherically symmetric thin shells, and provide 
a novel approach to the linearized stability analysis around a static solution. In
Section \ref{secIII}, we provide specific examples and consider a stability analysis by 
applying the generic linearized stability formalism outlined in section \ref{secII}. Finally, in 
Section \ref{conclusion}, we shall draw some general  conclusions.
Throughout this work, we adopt the sign conventions of Misner--Thorne--Wheeler \cite{MTWbook}, with $c=G=1$.

\section{General formalism}\label{secII}

To set the stage, consider two distinct spacetime manifolds, an \emph{exterior} ${\cal M_+}$, and an \emph{interior} ${\cal M_-}$, that are eventually to be joined together across some surface layer $\Sigma$. Let the two bulk spacetimes have 
metrics given by $g_{\mu \nu}^+(x^{\mu}_+)$ and $g_{\mu \nu}^-(x^{\mu}_-)$, in terms of 
independently defined coordinate systems $x^{\mu}_+$ and $x^{\mu}_-$. In particular, consider two generic static spherically symmetric spacetimes given by the following 
line elements:
\begin{eqnarray}
\fl
ds^2 = - e^{2\Phi_{\pm}(r_{\pm})}\left[1-\frac{b_{\pm}(r_{\pm})}{r_{\pm}}\right] dt_{\pm}^2 + 
\left[1-\frac{b_{\pm}(r_{\pm})}{r_{\pm}}\right]^{-1}\,dr_{\pm}^2 + r_{\pm}^2 d\Omega_{\pm}^{2}.
\label{generalmetric}
\end{eqnarray}
We take $+$ to refer to the exterior geometry and $-$ to refer to the interior geometry. For simplicity we assume that the exterior geometry is asymptotically flat and define
\begin{equation}
R_- = \max\{r: b_+(r)=r\} .
\end{equation}
(More generally, if the exterior geometry is not asymptotically flat, the maximum should be taken over whatever black hole horizons are present, but excluding the cosmological horizons.)
Similarly we assume that the interior geometry is regular at the origin ($r=0$) and define
\begin{equation}
R_+ = \min\{r: b_-(r)=r\} .
\end{equation}
Note the (at first glance) counter-intuitive placement of the $\pm$ on the quantities $R_\pm$: 
The conventions are chosen so that $R_+$ is the furthest outwards one can extend the interior 
geometry before hitting a horizon, whereas $R_-$ is the furthest inwards one can extend the 
exterior geometry before hitting a horizon.  

Since the whole point of a gravastar model is to avoid horizon formation we certainly desire 
$R_-< R_+$. In particular, if the thin-shell transition layer $\Sigma$ is located at 
$r=a(\tau)$, then to avoid horizon formation we demand
\begin{equation}
R_- < a(\tau) < R_+. 
\end{equation}
The key issue of central interest in this article is the dynamics of this surface layer.

\subsection{Bulk Einstein equations}

Using the Einstein field equation, $G_{{\mu}{\nu}}=8\pi \,T_{{\mu}{\nu}}$ (with $c=G=1$), the 
(orthonormal) stress-energy tensor components in the bulk are given by
\begin{eqnarray}
\fl
\rho(r)&=&\frac{1}{8\pi r^{2}}b',\label{rho}\\
\fl
p_{r}(r)&=&-\frac{1}{8\pi r^{2}}\left[2\Phi '(b-r)+b'\right],\label{pr}\\
\fl
p_{t}(r)&=&-\frac{1}{16\pi r^{2}}[(-b+3rb'-2r)\Phi '+2r(b-r)(\Phi ')^{2}+2r(b-r)\Phi ''
+b''r]\,,
\label{pt}
\end{eqnarray}
where the prime denotes a derivative with respect to the radial coordinate. Here $\rho(r)$ is 
the energy density, $p_r(r)$ is the radial pressure, and $p_t(r)$ is the lateral pressure 
measured in the orthogonal direction to the radial direction. The $\pm$ subscripts were 
(temporarily) dropped so as not to overload the notation. Note that in obtaining the individual field equations (\ref{rho})-(\ref{pt}), instead of using an orthonormal basis in the Einstein field equation one could simply, (because of the diagonal form of the metric in the current situation), consider the mixed tensor components, \emph{i.e.}, $G_{\mu}{}^{\nu}=8\pi \,T_{\mu}{}^{\nu}$. 

\subsection{Null energy condition}

Consider the null energy condition (NEC): $T_{\mu\nu}\,k^\mu\,k^\nu 
\geq 0$, where $T_{\mu\nu}$ is the stress-energy tensor and $k^{\mu}$ any null vector. Then along the 
radial direction, with $k^{\hat{\mu}}=(1,\pm 1,0,0)$ in the orthonormal frame where 
$T_{\hat{\mu}\hat{\nu}}={\rm diag}[\rho(r),p_r(r),p_t(r),p_t(r)]$, we have the particularly simple condition
\begin{equation}
 T_{\hat{\mu}\hat{\nu}}\,k^{\hat{\mu}}\,k^{\hat{\nu}}=\rho(r)+p_r(r)
 =\frac{(r-b)\Phi^{'}}{4\pi r^{2}}\geq 0.    \label{generalNEC}
\end{equation}
By hypothesis $r>b(r)$ in both the interior and exterior regions of the gravastar, so the radial NEC 
reduces to $\Phi'(r)>0$. The NEC in the transverse direction, $\rho+p_t\geq 0$, does not have any 
direct simple interpretation in terms of the metric components. In most gravastar models the NEC is taken to be satisfied, though the status of the NEC as fundamental physics is quite dubious~\cite{EC-violations}.

\subsection{Transition layer}

The interior and exterior manifolds are bounded by isometric hypersurfaces $\Sigma_+$ and $\Sigma_-$,  with induced 
metrics $g_{ij}^+$ and $g_{ij}^-$. By assumption
$g_{ij}^+(\xi)=g_{ij}^-(\xi)=g_{ij}(\xi)$, with natural hypersurface coordinates $\xi^i=(\tau, \theta, \phi)$. A single manifold ${\cal M}$ is obtained by gluing together ${\cal
M_+}$ and ${\cal M_-}$ at their boundaries. So ${\cal M}={\cal M_+}\cup {\cal M_-}$, with the 
natural identification of the boundaries $\Sigma=\Sigma_+=\Sigma_-$. The intrinsic metric on $\Sigma$ is
\begin{equation}
ds^2_{\Sigma}=-d\tau^2 + a(\tau)^2 \,(d\theta ^2+\sin
^2{\theta}\,d\phi^2).
\end{equation}
The position of the junction surface is given by $x^{\mu}(\tau,\theta,
\phi)=(t(\tau),a(\tau),\theta,\phi)$, and the respective $4$-velocities (as measured in the 
static coordinate systems on the two sides of the junction) are
\begin{eqnarray}
U^{\mu}_{\pm}=
\left(\frac{e^{-\Phi_{\pm}(a)}\sqrt{1-\frac{b_{\pm}(a)}{a}+\dot{a}^{2}}}{1-\frac{b_{\pm}(a)}
{a}},\;
\dot{a},0,0 \right).
\end{eqnarray}
The overdot denotes a derivative with respect to $\tau$,  the proper time of an observer 
comoving with the junction surface. The Israel formalism requires that the normals point from 
${\cal M_-}$ to ${\cal M_+}$ \cite{junction_formalism}. The unit normals to the junction 
surface are
\begin{eqnarray}
n^{\mu}_{\pm}= \left(
\frac{e^{-\Phi_{\pm}(a)}}{1-\frac{b_{\pm}(a)}{a}}\;\dot{a},
\sqrt{1-\frac{b_{\pm}(a)}{a}+\dot{a}^2},0,0 
\right) \label{normal}
\,.
\end{eqnarray}
In view of the spherical symmetry these results can easily be deduced from the contractions $U^{\mu}U_{\mu}=-1$, 
$U^{\mu}n_{\mu}=0$, and $n^{\mu}n_{\mu}=+1$.
The extrinsic curvature, or the second fundamental form, is defined as $K_{ij}=n_{\mu;
\nu}e^{\mu}_{(i)}e^{\nu}_{(j)}$. Differentiating $n_{\mu}e^{\mu}_{(i)}=0$ with respect to 
$\xi^j$, we have
\begin{equation}
n_{\mu}\frac{\partial ^2 x^{\mu}}{\partial \xi^i \, \partial \xi^j}=
-n_{\mu,\nu}\, \frac{\partial x^{\mu}}{\partial \xi^i}\frac{\partial x^{\nu}}{\partial \xi^j},
\end{equation}
so that general the extrinsic curvature is given by
\begin{eqnarray}
\label{extrinsiccurv}
K_{ij}^{\pm}=-n_{\mu} \left(\frac{\partial ^2 x^{\mu}}{\partial
\xi ^{i}\,\partial \xi ^{j}}+\Gamma ^{\mu \pm}_{\;\;\alpha
\beta}\;\frac{\partial x^{\alpha}}{\partial \xi ^{i}} \,
\frac{\partial x^{\beta}}{\partial \xi ^{j}} \right) \,.
\end{eqnarray}
For a thin shell $K_{ij}$ is not continuous across $\Sigma$. For 
notational convenience, the discontinuity in the second fundamental form is defined as 
$\kappa_{ij}=K_{ij}^{+}-K_{ij}^{-}$.
The non-trivial components of the extrinsic curvature can 
easily be computed to be
\begin{eqnarray}
K^{\theta \;\pm }_{\;\;\theta}&=&
\frac{1}{a}\,\sqrt{1-\frac{b_{\pm}(a)}{a}+\dot{a}^2}\;,
\label{genKplustheta}
\\
K^{\tau\;\pm}_{\;\;\tau}&=&
\left\{\frac{\ddot a+\frac{b_{\pm}(a)-b'_{\pm}(a)a}{2a^2}}{\sqrt{1-\frac{b_{\pm}(a)}{a}+\dot{a}^{2}}} 
+ \Phi'_{\pm}(a) \sqrt{1-\frac{b_{\pm}(a)}{a}+\dot{a}^{2}} \right\}  \,, 
\label{genKminustautau}
\end{eqnarray}
where the prime now denotes a derivative with respect to the coordinate $a$.
\begin{itemize}
\item 
Note that $K^{\theta \;\pm }_{\;\;\theta}$ is independent of the quantities $\Phi_\pm$. This is most easily verified by noting that in terms of the 
normal distance $\ell$ to the shell $\Sigma$ the extrinsic curvature can be written as $K_{ij} = 
{1\over2} \partial_\ell g_{ij} = {1\over2} n^\mu \partial_\mu g_{ij} =  {1\over2} n^r \partial_r 
g_{ij}$, where the last step relies on the fact that the bulk spacetimes are static. Then since 
$g_{\theta\theta}=r^2$, differentiating and setting $r\to a$ we have $K_{\theta\theta} = a \; n^r$. 
Thus 
\begin{equation}
K^{\theta \;\pm }_{\;\;\theta} = {n^r\over a},
\end{equation}
which is a particularly simple formula in terms of the radial component of the normal vector, and 
which easily lets us verify \eqref{genKplustheta}.

\item
For $K_{\tau\tau}$ there is an argument (easily extendable to the present context) in 
reference~\cite{Visser} (see especially pages 181--183) to the effect that
\begin{eqnarray}
\fl 
K^{\tau\;\pm}_{\;\;\tau} &=& \g_\pm = \hbox{(magnitude of the physical 4-acceleration of the transition layer)}.
\nonumber\\
\fl &&
\end{eqnarray}
This gives a clear physical interpretation to $K^{\tau\;\pm}_{\;\;\tau}$ and rapidly allows one to 
verify \eqref{genKminustautau}.

\item
There is also an important differential relationship between these extrinsic curvature components 
\begin{equation}
{\d\over\d\tau}\left\{ a \; e^{\Phi_\pm} \; K^{\theta \;\pm }_{\;\;\theta} \right\} = e^{\Phi_\pm} 
\; K^{\tau\;\pm}_{\;\;\tau} \; \dot a.
\label{E:differential}
\end{equation}
The most direct way to verify this is to simply differentiate, using \eqref{genKplustheta} and 
\eqref{genKminustautau} above. Geometrically, the existence of these relations between the extrinsic 
curvature components is ultimately due to the fact that the bulk spacetimes have been chosen to be 
static. By noting that
\begin{equation}
{\d\over \d a}\left({1\over2} \dot a^2\right) = \left({\d\over \d a} \dot a\right)  \dot a = \ddot a,
\label{E:trick}
\end{equation}
we can also write this differential relation as
\begin{equation}
{\d\over\d a}\left\{ a \; e^{\Phi_\pm} \; K^{\theta \;\pm }_{\;\;\theta} \right\} = e^{\Phi_\pm} 
\; K^{\tau\;\pm}_{\;\;\tau}.
\end{equation}
\end{itemize}

\subsection{Lanczos equations: Surface stress-energy}

The Lanczos equations follow from the Einstein equations applied to the hypersurface joining 
the bulk spacetimes, and are given by
\begin{equation}
S^{i}_{\;j}=-\frac{1}{8\pi}\,(\kappa ^{i}_{\;j}-\delta^{i}_{\;j}\;\kappa ^{k}_{\;k})  \,.
\end{equation}
Here $S^{i}_{\;j}$ is the surface stress-energy tensor on $\Sigma$. In particular, because of 
spherical symmetry considerable simplifications occur, namely $\kappa ^{i}_{\;j}={\rm diag}
\left(\kappa ^{\tau}_{\;\tau},\kappa ^{\theta}_{\;\theta},\kappa^{\theta}_{\;\theta}\right)$. 
The surface stress-energy tensor may be written in terms of the surface energy density, 
$\sigma$, and the surface pressure, $\P$, as $S^{i}_{\;j}={\rm diag}(-\sigma,\P,\P)$. The 
Lanczos equations then reduce to
\begin{equation}
\sigma =-{\kappa ^{\theta}_{\;\theta}\over4\pi};
\qquad
\P ={\kappa ^{\tau}_{\;\tau}+\kappa^{\theta}_{\;\theta}\over8\pi};
\qquad
\sigma+2\P ={\kappa ^{\tau}_{\;\tau}\over4\pi}\,.
\label{lanczos}
\end{equation}
From equations~(\ref{genKplustheta})--(\ref{genKminustautau}), we see that:
\begin{eqnarray}
\fl
\sigma&=&-\frac{1}{4\pi a}\left[
 \sqrt{1-\frac{b_{+}(a)}{a}+\dot{a}^{2}}
-\sqrt{1-\frac{b_{-}(a)}{a}+\dot{a}^{2}}
\right],
\label{gen-surfenergy2}
\\
\fl
\P&=&\frac{1}{8\pi a}\left[
\frac{1+\dot{a}^2+a\ddot{a}-\frac{b_+(a)+ab'_+(a)}{2a}}{\sqrt{1-\frac{b_{+}(a)}{a}+\dot{a}^{2}}}     
+
\sqrt{1-\frac{b_{+}(a)}{a}+\dot{a}^{2}} \; a\Phi'_{+}(a)
\right. \nonumber\\
\fl
&&
\qquad 
\left. -
\frac{1+\dot{a}^2+a\ddot{a}-\frac{b_-(a)+ab'_-(a)}{2a}}{\sqrt{1-\frac{b_{-}(a)}{a}+\dot{a}^{2}}}
-
\sqrt{1-\frac{b_{-}(a)}{a}+\dot{a}^{2}} \; a\Phi'_{-}(a)
\right],\label{gen-surfpressure2}
\end{eqnarray}
and finally
\begin{eqnarray}
\fl
\sigma+2\P&=& {[\g]\over4\pi} = \frac{1}{4\pi}\left[
\frac{\ddot{a}+\frac{b_+(a)-ab'_+(a)}{2a^2}}{\sqrt{1-\frac{b_{+}(a)}{a}+\dot{a}^{2}}}     
+
\sqrt{1-\frac{b_{+}(a)}{a}+\dot{a}^{2}} \; \Phi'_{+}(a)
\right. \nonumber\\
\fl 
&&
\qquad \qquad
\left. -
\frac{\ddot{a}+\frac{b_-(a)-ab'_-(a)}{2a^2}}{\sqrt{1-\frac{b_{-}(a)}{a}+\dot{a}^{2}}}
-
\sqrt{1-\frac{b_{-}(a)}{a}+\dot{a}^{2}} \; \Phi'_{-}(a)
\right].\label{s2P}
\end{eqnarray}
Note that $\sigma + 2 \P$ has a particularly simple physical interpretation in terms of  
$[\g]$,  the discontinuity in 4-acceleration. 
(This is ultimately related to the fact that the quantity $\sigma+2\P$ for a thin shell has 
properties remarkably similar to those of the quantity $\rho+3p$ for a bulk spacetime.) 
Furthermore the 
surface energy density $\sigma$  is independent of the quantities $\Phi_\pm$. 
The surface mass of the thin shell is given by $m_s=4\pi a^2\sigma$. 

Independent of the state of motion of the thin shell,  we have $\sigma(a) > 0$ 
whenever $b_+(a)> b_-(a)$, and  $\sigma(a) < 0$ whenever $b_+(a)< b_-(a)$. The situation where 
$\sigma=0$ corresponds to $b_-(a)=b_+(a)$, and is precisely the case where all the 
discontinuities are concentrated in $K^{\tau}_{\;\tau}$ while $K^{\theta}_{\;\theta}$ is 
continuous. This phenomenon, the vanishing of $\sigma$ at certain specific shell radii given 
by $b_-(a)=b_+(a)$, is generic to gravastars but (because of a few key sign flips) cannot 
occur for the thin-shell traversable wormholes considered in 
\cite{GLV, Visser, thinshellWH}. This is perhaps the most obvious of many properties 
differentiating gravastars from the thin-shell traversable wormhole case, although wormhole 
geometries surrounded by thin shells, similar to the cases explored in this work, have also been 
analyzed in the literature \cite{thinshellWH2,gr-qc/0409018}.

\subsection{Static gravastars}\label{secII_static}

Assume, for the sake of discussion, a static solution at some $a_0 \in(R_-,R_+)$. Then
\begin{eqnarray}
\sigma(a_0)&=&-\frac{1}{4\pi a_0}\left[
 \sqrt{1-\frac{b_{+}(a_0)}{a_0}}
-\sqrt{1-\frac{b_{-}(a_0)}{a_0}}
\right],
\label{gen-surfenergy2a}
\\
\P(a_0)&=&\frac{1}{8\pi a_0}\left[
\frac{1-\frac{b_+(a_0)+a_0b'_+(a_0)}{2a_0}}{\sqrt{1-\frac{b_{+}(a_0)}{a_0}}}     
+
\sqrt{1-\frac{b_{+}(a_0)}{a_0}} \; a_0\Phi'_{+}(a_0)
\right. \nonumber\\
&&
\qquad 
\left. -
\frac{1-\frac{b_-(a_0)+a_0b'_-(a_0)}{2a_0}}{\sqrt{1-\frac{b_{-}(a_0)}{a_0}}}
-
\sqrt{1-\frac{b_{-}(a_0)}{a_0}} \; a_0\Phi'_{-}(a_0)
\right],\label{gen-surfpressure2a}
\end{eqnarray}
and finally
\begin{eqnarray}
\sigma(a_0)+2\P(a_0)&=& {[\g_0]\over4\pi} = \frac{1}{4\pi}\left[
\frac{\frac{b_+(a)-ab'_+(a)}{2a^2}}{\sqrt{1-\frac{b_{+}(a)}{a}}}     
+
\sqrt{1-\frac{b_{+}(a)}{a}} \; \Phi'_{+}(a)
\right. \nonumber\\
&&
\qquad\qquad
\left. -
\frac{\frac{b_-(a)-ab'_-(a)}{2a^2}}{\sqrt{1-\frac{b_{-}(a)}{a}}}
-
\sqrt{1-\frac{b_{-}(a)}{a}} \; \Phi'_{-}(a)
\right].\label{s2P}
\nonumber\\
&&
\end{eqnarray}
(See also equation (32) of~\cite{Abreu}.)
Now taking $a_0\to R_-$,  we have
\begin{equation}
\sigma(R_-)=+\frac{1}{4\pi R_-}
\sqrt{1-\frac{b_{-}(R_-)}{R_-}}  > 0.
\end{equation}
However taking $a_0\to R_+$ we have
\begin{equation}
\sigma(R_+)=-\frac{1}{4\pi R_+}
\sqrt{1-\frac{b_{+}(R_+)}{R_+}}  < 0.
\end{equation}
That is:
\begin{equation}
\sigma(R_\pm)= \mp\frac{1}{4\pi R_\pm}
\sqrt{1-\frac{b_{\pm}(R_\pm)}{R_\pm}}.
\end{equation}
Applying the mean value theorem, for gravastars there will \emph{always} be some $R_0 
\in(R_-,R_+)$, possibly many such $R_0$,  such that $\sigma(R_0)=0$. (This phenomenon 
\emph{cannot} occur for the thin-shell traversable wormholes considered in 
\cite{GLV, Visser, thinshellWH}, because of key sign flips --- for thin-shell traversable wormholes we 
always have $\sigma < 0$.) The $R_0$ such that $\sigma(R_0)=0$ is clearly a special place for 
gravastars.  Explicitly this occurs when
\begin{equation}
b_+(R_0) = b_-(R_0),
\end{equation}
and in fact for 
\begin{equation}
R_-< b_+(R_0) = b_-(R_0) < R_+.
\end{equation}
At this special point the discontinuities are concentrated in $K^{\tau}_{\;\tau}$ while $K^{\theta}_{\;\theta}$ is continuous. That is
\begin{equation}
\fl
\P(R_0)=\frac{1}{16\pi R_0}\left[
\frac{b'_-(R_0)-b'_+(R_0)}{\sqrt{1-\frac{b_\pm(R_0)}{R_0}}}     
+
2\sqrt{1-\frac{b_\pm(R_0)}{R_0}} \; R_0[\Phi'_{+}(R_0)-\Phi'_-(R_0)]
\right] \,,
\end{equation}
so that
\begin{equation}
\P(R_0)= {[K^{\tau}_{\;\tau}]\over8\pi}={[\g_0]\over8\pi}.
\end{equation}
Even though $\sigma=0$, one needs $\P\neq 0$ because of the non-zero 4-acceleration $\g_0$.

If one also demands that the surface pressure at $R_0$ also be zero, $\P(R_0)=0$, one must impose the additional condition
\begin{equation}
b'_+-b'_- = 2\left( R_0-b_\pm \right)\left( \Phi'_+ - \Phi'_- \right),
\end{equation}
which is equivalent to
\begin{equation}
\left.\left(  e^{2\Phi_+} [1-b_+/a_0] \right)'\right|_{R_0}= \left.\left(  e^{2\Phi_-} [1-b_-/a_0] \right)'\right|_{R_0}\,.
\end{equation}
If $\Phi'_\pm(R_0)$, then not only does $b_+(R_0)=b_-(R_0)$ but also $b_+'(R_0)=b'_-(R_0)$ --- these two conditions give a zero pressure and zero density shell. 
More specifically, in terms of standard  nomenclature, if the surface stress-energy terms are zero, the 
junction is denoted as a boundary surface; if surface stress terms are present, the junction 
is called a thin shell.

\subsection{Surface stress estimates}\label{secII_srtress_estimates}

It is interesting to obtain some estimates of the surface stresses. For this purpose, consider for simplicity an exterior geometry that is Schwarzschild-de Sitter spacetime, so that 
\begin{equation}
b_+(r)= 2M + \frac{\Lambda}{3}r^3 \,, \qquad \Phi_+(r)=0 \,.
\end{equation}
An important quantity that will be play a fundamental role throughout this paper is the surface mass of the thin shell, which is given by $m_s=4\pi a^2\sigma$. For the exterior Schwarzschild-de Sitter spacetime, and considering an arbitrary interior geometry, the surface mass of the static thin shell is given by
\begin{eqnarray}\label{shellmass}
m_{s}(a_0)=a_0\left(\sqrt{1-\frac{b_-(a_0)}{a_0}}-\sqrt{1-\frac{2M}{a_0}-
\frac{\Lambda}{3}a_0^2} \, \right)  .
\end{eqnarray}
Note that one may interpret $M$ as the total mass of the system, as measured in an asymptotic region of the spacetime. Solving for $M$, we have
\begin{equation}\label{totalmass}
M=\frac{b_-(a_0)}{2}+m_{s}(a_0)\left(\sqrt{1-\frac{b_-(a_0)}{a_0}}-\frac{m_{s}(a_0)}{2a_0}\right)-\frac{\Lambda}{6}a_0^3   \,.
\end{equation}
For the Schwarzschild--de~Sitter spacetime $\Lambda >0$. For the range $0<9\Lambda M^2<1$, the factor $g_{rr}^{-1}=-g_{tt}=(1-2M/r-\Lambda r^2/3)$ possesses two positive real roots, $r_b$ and $r_c$, corresponding to the black hole and the cosmological event horizons:
\begin{eqnarray}
r_b&=&2 \Lambda ^{-1/2} \, \cos(\alpha/3)    \label{root1}  \,, \\
r_c&=&2 \Lambda ^{-1/2} \, \cos(\alpha/3+4\pi/3)    \label{root2}
\,.
\end{eqnarray}
Here $\cos \alpha \equiv -3M \Lambda^{1/2}$, with $\pi < \alpha
<3\pi/2$. In this domain we have $2M<r_b<3M$ and $r_c>3M$ \cite{gr-qc/0409018}.

Defining suitable dimensionless parameters, equations (\ref{gen-surfenergy2a})--(\ref{gen-surfpressure2a}) take, for the Schwarzschild-de
Sitter solution,  the form
\begin{eqnarray}
\fl
\mu(a_0)&=&x \left(\sqrt{1-x\,\bar{b}(x)}-\sqrt{1-x-
\frac{4\beta}{27x^2}} \; \right)
    \label{mu}   ,\\
\fl 
\Pi(a_0)&=&x \left(\frac{1-\frac{x}{2}
-\frac{8\beta}{27x^2}}{\sqrt{1-x-\frac{4\beta}{27x^2}}}
-\frac{1-\frac{1}{2}\left[ x\bar{b}(x)+b'_- \right] }{\sqrt{1-x\bar{b}(x)}}
-\zeta_- \, \sqrt{1-x\bar{b}(x)} \, \right)
    \label{Pi}    .
\end{eqnarray}
Here we define: $x=2M/a$, $\beta=9\Lambda M^2$,
$\bar{b}(x)=b(a)/(2M)$, $\mu=8\pi M\sigma$, and set $\Pi=16\pi M{\cal
P}$. In the analysis that follows we shall assume that $M$ is
positive, $M>0$.

\begin{figure}[!htb]
  \includegraphics[width=3 in]{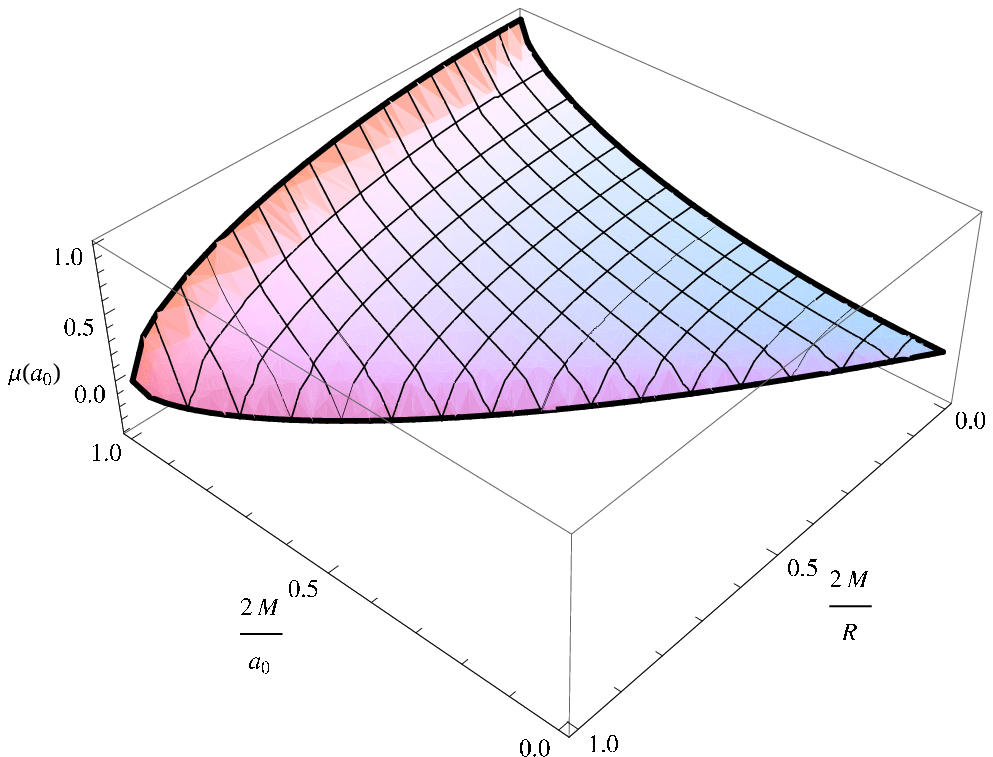}
  \includegraphics[width=3 in]{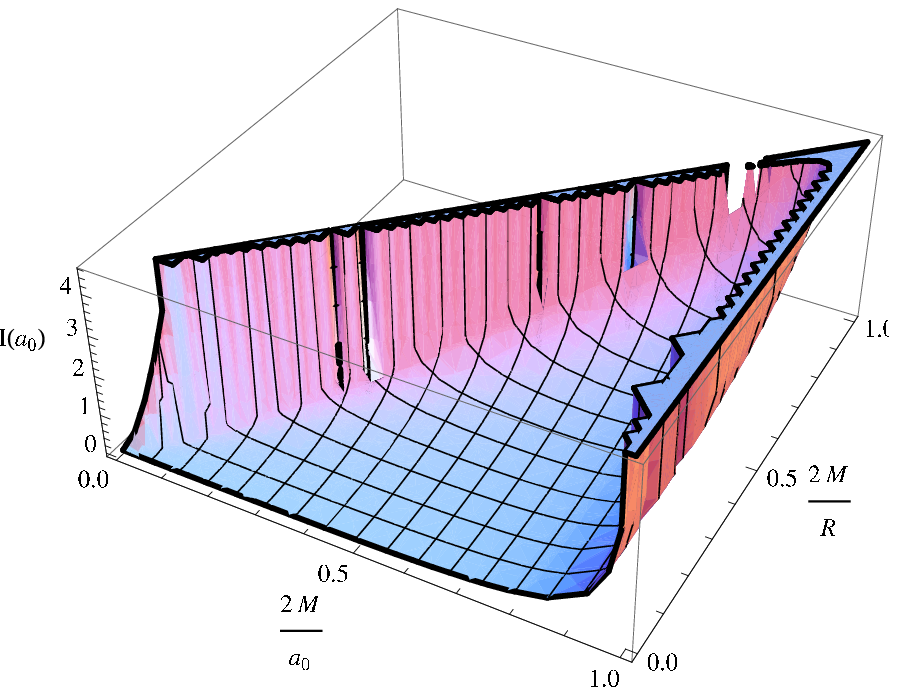}
  \caption{Schwarzschild--de~Sitter gravastar: These plots depict the qualitative behaviour of the surface stresses of a gravastar 
  with interior de Sitter and exterior Schwarzschild spacetimes.  We have 
  considered the dimensionless parameters, $x=2M/a_0$ and $y=2M/R$, and 
  $\mu(a_0)=8\pi M\sigma(a_0)$ and $\Pi(a_0)=16\pi M\P(a_0)$. The left plot represents the 
  surface energy density, $\sigma$. Note that $\sigma$ is positive in the range $2M<a_0< 
  (2M R^2)^{1/3}$, and negative in the range $(2M R^2)^{1/3} < a_0 < R$.
  The right plot represents the surface pressure, $\P$. The latter surface pressure diverges 
  as $a_0$ approaches $2M$ ($R$) from the right (left). See the text for more details.}
  \label{plotsSchdS1}
\end{figure}

As a specific application, consider the standard gravastar picture, which consists of a Schwarzschild exterior geometry and an interior de Sitter spacetime. Thus the surface stresses, equations (\ref{mu})--(\ref{Pi}), are obtained by setting $\Lambda=0$, (i.e., $\beta=0$), while $b_-(a_0) = a_0^3/R^2$ and $\Phi_-(a_0)=0$. Here we have defined $R^2=3/\Lambda_-$, for simplicity. For this case the total mass of the system $M$ is given by
\begin{equation}
M=\frac{a_0^2}{2R^2}+m_s(a_0)\left(\sqrt{1-\frac{a_0^2}{R^2}}-\frac{m_s}{2a_0} \right)\,.
\end{equation}
Using the dimensionless parameters, $x=2M/a_0$ and $y=2M/R$, while $\mu(a_0)=8\pi M\sigma(a_0)$ and $\Pi(a_0)=16\pi M\P(a_0)$, equations  (\ref{mu})--(\ref{Pi}), take the form
\begin{eqnarray}
\mu(a_0) = -x\left( \sqrt{1-x} - \sqrt{1-\frac{y^2}{x^2}} \right) \,, 
  \\
\Pi(a_0) = x\left( \frac{1-\frac{x}{2}}{\sqrt{1-x}} - \frac{1-2\frac{y^2}{x^2}}{\sqrt{1-\frac{y^2}{x^2}}} \right) \,.
\end{eqnarray}
The qualitative behaviour is depicted in figure \ref{plotsSchdS1}. The left plot represents the 
surface energy density, $\sigma$. Note that $\sigma$ is positive in the range $2M<a_0< 
(2M R^2)^{1/3}$, and negative for $(2M R^2)^{1/3} < a_0 < R$. The right plot represents 
the surface pressure, $\P$. The latter diverges as $a_0$ approaches $2M$ ($R$) from the right 
(left). A detailed stability analysis of the thin shell for this specific configuration will be presented in section \ref{specific_example1}.

\subsection{Conservation identity}

The first contracted Gauss--Codazzi equation\footnote{Sometimes called simply the Gauss equation,  or in 
general relativity more often referred to as the ``Hamiltonian constraint''.} is
\begin{eqnarray}
G_{\mu \nu}\;n^{\mu}\,n^{\nu}=\frac{1}{2}\,(K^2-K_{ij}K^{ij}-\,^3R)\,.
    \label{1Gauss}
\end{eqnarray}
The second contracted Gauss--Codazzi equation\footnote{Sometimes called simply the Codazzi or 
the Codazzi--Mainardi equation, or in general relativity more often referred to as the ``ADM 
constraint'' or ``momentum constraint''.} is
\begin{eqnarray}
G_{\mu \nu}e^{\mu}_{(i)}n^{\nu}=K^j_{i|j}-K,_{i}\,.
    \label{2Gauss}
\end{eqnarray}
Together with the Lanczos equations this provides the conservation identity
\begin{eqnarray}\label{conservation}
S^{i}_{\;j|i}=\left[T_{\mu \nu}\; e^{\mu}_{\;(j)}n^{\nu}\right]^+_-\,,
\end{eqnarray}
where the convention $\left[X \right]^+_-\equiv X^+|_{\Sigma}-X^-|_{\Sigma}$ is used.
When interpreting this conservation identity, consider first the momentum
flux defined by
\begin{equation}
\fl 
\left[T_{\mu\nu}\; e^{\mu}_{\;(\tau)}\,n^{\nu}\right]^+_-
=\left[T_{\mu\nu}\; U^{\mu}\,n^{\nu}\right]^+_-
=\left[
\left(T_{\hat{t}\hat{t}}+T_{\hat{r}\hat{r}}\right) 
\,\frac{\dot{a}\sqrt{1-\frac{b(a)}{a}+\dot{a}^{2}}}{1-\frac{b(a)}{a}} \;
\right]^+_-\,,   
\label{flux}
\end{equation}
where $T_{\hat{t}\hat{t}}$ and $T_{\hat{r}\hat{r}}$ are the bulk stress-energy tensor 
components given in an orthonormal basis. This flux term corresponds to the net discontinuity 
in the (bulk) momentum flux $F_\mu=T_{\mu\nu}\,U^\nu$ which impinges on the shell. Applying 
the (bulk) Einstein equations we see
\begin{equation}
\fl
\left[T_{\mu\nu}\; e^{\mu}_{\;(\tau)}\,n^{\nu}\right]^+_-
=\frac{\dot a}{4\pi a}\, \left[
\Phi_+'(a)\sqrt{1-\frac{b_+(a)}{a}+\dot{a}^{2}}
-
\Phi_-'(a)\sqrt{1-\frac{b_-(a)}{a}+\dot{a}^{2}}
\right]\,.
\end{equation}
It is useful to define the quantity
\begin{equation}
\Xi 
=\frac{1}{4\pi a}\, \left[
\Phi_+'(a)\sqrt{1-\frac{b_+(a)}{a}+\dot{a}^{2}}
- 
\Phi_-'(a)\sqrt{1-\frac{b_-(a)}{a}+\dot{a}^{2}}
\right]\,,
\end{equation}
and to let $A=4\pi a^2$ be the surface area of the thin shell.
Then in the general case, the conservation identity provides the following relationship
\begin{equation}
\frac{d\sigma}{d\tau}+(\sigma+\P)\,{1\over A} \frac{dA}{d\tau}=\Xi \, \dot{a}\,,
\label{E:conservation1}
\end{equation}
or equivalently
\begin{equation}
\frac{d(\sigma A)}{d\tau}+\P\,\frac{dA}{d\tau}=\Xi \,A \, \dot{a}\,.
\label{E:conservation2}
\end{equation}
The first term represents the variation of the internal energy of the shell, the second term 
is the work done by the shell's internal force, and the third term represents the work done by 
the external forces.  Once could also brute force verify this equation by explicitly 
differentiating \eqref{gen-surfenergy2} using \eqref{gen-surfpressure2} and the relations 
\eqref{E:differential}. If we assume that the equations of motion can be integrated to 
determine the surface energy density as a function of radius $a$, that is, assuming the 
existence of a suitable function $\sigma(a)$, then the conservation equation can be written as
\begin{equation}
\sigma'=-\frac{2}{a}\,(\sigma +\P)+\Xi\,,    
\label{cons-equation}
\end{equation}
where $ \sigma'=d\sigma /da$. We shall carefully analyze the integrability conditions for 
$\sigma(a)$ in the next sub-section. For now, note that the flux term (external force term) $\Xi$ is 
automatically zero whenever $\Phi_\pm=0$; this is actually a quite common occurrence, for instance 
in either Schwarzschild or Reissner--Nordstr\"om geometries, or more generally whenever $\rho+p_r=0$, 
so it is very easy for one to be mislead by those special cases. In particular, in situations of 
vanishing flux $\Xi=0$ one obtains the so-called  ``transparency condition'', $\left[G_{\mu
\nu}\; U^{\mu}\,n^{\nu}\right]^+_-=0$, see \cite{Ishak}. The conservation identity, 
equation~(\ref{conservation}), then reduces to  the simple relationship $\dot{\sigma}=-2\,(\sigma +{\cal 
P}) \dot{a}/a$. But in general the ``transparency condition'' does not hold, and one needs the full 
version of the conservation equation as given in equation~\eqref{E:conservation2}.

\subsection{Integrability of the surface energy density}

When does it make sense to assert the existence of a function $\sigma(a)$? Let us start with 
the situation in the absence of external forces (we will rapidly generalize this) where the 
conservation equation, 
\begin{equation}
\dot{\sigma}=-2\,(\sigma +\P) \dot{a}/a \,,
\end{equation}
can easily be rearranged to
\begin{equation}
{\dot\sigma\over\sigma +\P} = -2 {\dot a\over a}.
\end{equation}
Assuming a barotropic equation of state $\P(\sigma)$ for the matter in the gravastar transition layer, 
this can be integrated to yield
\begin{equation}
\int_{\sigma_0}^\sigma {\d\bar\sigma\over\bar\sigma +\P(\bar\sigma) } = -2 \int_{a_0}^a 
{\d\bar a\over \bar a} = -2 \ln(a/a_0).
\end{equation}
This implies that $a$ can be given as some function $a(\sigma)$ of $\sigma$, and by the inverse 
function theorem implies over suitable domains the existence of a function $\sigma(a)$. Now this 
barotropic equation of state is a rather strong assumption, albeit one that is very often implicitly 
made when dealing with thin-shell gravastar (or thin-shell wormholes, or other thin-shell objects).  
As a first generalization, consider what happens if the surface pressure generalized is to be of the form $\P(a,
\sigma)$, which is not barotropic. Then the conservation equation can be rearranged to be
\begin{equation}
\sigma'= - {2[\sigma +\P(a,\sigma)]\over a}.
\end{equation}
This is a first-order (albeit nonlinear and non-autonomous) ordinary differential equation, which at 
least locally will have solutions $\sigma(a)$. There is no particular reason to be concerned about 
the question of global solutions to this ODE, since in applications one is most typically dealing 
with linearization around a static solution.

If we now switch on external forces, one way of guaranteeing integrability would be to demand that the external forces are of the form $\Xi(a,\sigma)$, since then the conservation equation would read
\begin{equation}
\sigma'= - {2[\sigma +\P(a,\sigma)]\over a} + \Xi(a,\sigma),
\end{equation}
which is again a first-order albeit nonlinear and non-autonomous ordinary differential equation. But 
how general is this $\Xi = \Xi(a,\sigma)$ assumption? There are at least two nontrivial situations where this 
definitely holds:
\begin{itemize}
\item If $\Phi_+(a) = \Phi_-(a) = \Phi(a)$, then $\Xi = - \Phi'(a) \; \sigma$, which is explicitly 
of the required form.
\item If $b_+(a)=b_-(a)=b(a)$, \emph{but the $\Phi_\pm$ are unequal}, then  $\sigma\equiv 0$ regardless of the location and state of motion of the transition layer. Furthermore $\Xi \equiv 2\P/a$. 
(This would make for a somewhat unusual gravastar.)
\item
If both $b_+(a)=b_-(a)=b(a)$ and $\Phi_+(a) = \Phi_-(a) = \Phi(a)$, then the situation is vacuous. There is then no discontinuity in extrinsic curvatures and the thin shell carries no stress-energy; so this in fact corresponds to a ``continuum'' gravastar with $b(r)/r < 1$ for all $r\in(0,\infty)$. 
\end{itemize}
But in general we will need a more complicated set of assumptions to assure integrability, and 
the consequent existence of a function $\sigma(a)$. A model that is always \emph{sufficient} 
(not necessary) to guarantee integrability is to view the exotic material in the transition layer as a 
two-fluid system, characterized by $\sigma_\pm$ and $\P_\pm$, with two (possibly independent) 
equations of state $\P_\pm(\sigma_\pm)$. Specifically, take
\begin{eqnarray}
\sigma_\pm &=&-\frac{1}{4\pi}\,(K_\pm)^{\theta}_{\;\theta} \,,
\\
\P_\pm &=&\frac{1}{8\pi}\left\{ (K_\pm)^{\tau}_{\;\tau}+(K_\pm)^{\theta}_{\;\theta}\right\} 
\,.
\end{eqnarray}
In view of the differential identities
\begin{equation}
{\d\over\d\tau}\left\{ a \; e^{\Phi_\pm} \; K^{\theta \;\pm }_{\;\;\theta} \right\} = e^{\Phi_\pm} 
\; K^{\tau\;\pm}_{\;\;\tau} \; \dot a,
\end{equation}
each of these two fluids is independently subject to
\begin{equation}
{\d\over\d\tau}\left\{ e^{\Phi_\pm} \; \sigma_\pm  \right\} = -{2e^{\Phi_\pm}\over a} \; 
\{\sigma_\pm + \P_\pm\} \; \dot a,
\end{equation}
which is equivalent to
\begin{equation}
\left\{ e^{\Phi_\pm} \; \sigma_\pm  \right\}' = -{2e^{\Phi_\pm}\over a} \; \{\sigma_\pm + {\cal 
P}_\pm\}.
\end{equation}
With two equations of state $\P_\pm(\sigma_\pm)$ these are two nonlinear first-order ordinary differential equations for 
$\sigma_\pm$.  These equations are integrable, implicitly defining functions $\sigma_\pm(a)$, at least locally. 
Once this is done we define
\begin{equation}
\sigma(a) = \sigma_+(a) -  \sigma_-(a),
\end{equation}
and
\begin{equation}
m_s(a) = 4\pi \sigma(a) \; a^2.
\end{equation}
While the argument is more complicated than one might have expected, the end result is easy to 
interpret: We can simply \emph{choose} $\sigma(a)$, or equivalently $m_s(a)$, as an arbitrarily 
specifiable function that encodes the (otherwise unknown) physics of the specific form of
matter residing on the gravastar transition layer.

\subsection{Equation of motion}

To qualitatively analyze the stability of the gravastar, assuming integrability of the surface energy 
density, (that is, the existence of a function $\sigma(a)$), it is useful to rearrange 
equation~(\ref{gen-surfenergy2}) into the form
\begin{equation}
{1\over2} \dot{a}^2+V(a)=0  \,,
\end{equation}
where the potential $V(a)$ is given by\footnote[1]{This equation only valid for $\sigma\not\equiv0$ due to a divide-by-zero problem.  The $\sigma\equiv0$ case is a special one worth separate consideration:
\[
V(a)= {1\over2}\left\{ 1-{\bar b(a)\over a}  -\left({8\pi\P\over\Phi_+'-\Phi_-'}\right)^2 \right\}\,.
\]
}
\begin{equation}
V(a)= {1\over2}\left\{ 1-{\bar b(a)\over a} -\left[\frac{m_{s}(a)}{2a}\right]^2-\left[\frac{\Delta(a)}{m_{s}
(a)}\right]^2\right\}\,.
   \label{potential}
\end{equation}
Here $m_s(a)=4\pi a^2\,\sigma(a)$ is the mass of the thin shell. The quantities $\bar b(a)$ and 
$\Delta(a)$ are defined, for simplicity, as
\begin{eqnarray}
\bar b(a)&=&\frac{b_{+}(a)+b_{-}(a)}{2},\\
\Delta(a)&=&\frac{b_{+}(a)-b_{-}(a)}{2},
\end{eqnarray}
respectively. This gives the potential $V(a)$ as a function of the surface mass $m_s(a)$.
By differentiating with respect to $a$, (using \eqref{E:trick}), we see that  the equation of motion implies
\begin{equation}
\ddot a = -V'(a).
\end{equation}
It is sometimes useful to reverse the logic flow and determine the surface mass as a function of the 
potential. Following the techniques used in~\cite{gr-qc/0310107, GLV}, suitably modified for the present context, a brief calculation yields
\begin{equation}
m_s(a) = -a
\left[ 
  \sqrt{ 1- {b_+(a)\over a} - 2V(a)} -\sqrt{ 1- {b_-(a)\over a} - 2V(a)} 
\right],
\end{equation}
with the negative root now being necessary for compatibility with the Lanczos equations. Note the 
logic here --- assuming integrability of the surface energy density, if we want a specific $V(a)$ 
this tells us how much surface mass we need to put on the transition layer (as a function of $a$), 
which is implicitly making demands on the equation of state of the matter residing on the 
transition layer. In a completely analogous manner, the assumption of integrability of $\sigma(a)$ implies that 
after imposing the equation of motion for the shell one has
\begin{equation}
\sigma(a)=-\frac{1}{4\pi a}\left[\sqrt{1-\frac{b_{+}(a)}{a} - 2 V(a)}-\sqrt{1-\frac{b_{-}(a)}{a}- 2 V(a)}\right],
\label{gen-surfenergy2-onshell}
\end{equation}
while
\begin{eqnarray}
\fl
\P&=&\frac{1}{8\pi a}\left[
\frac{1-2V(a)-aV'(a)-\frac{b_+(a)+ab'_+(a)}{2a}}{\sqrt{1-\frac{b_{+}(a)}{a}-2V(a)}}     
+
\sqrt{1-\frac{b_{+}(a)}{a}-2V(a)} \; a\Phi'_{+}(a)
\right. \nonumber\\
\fl
&&
\qquad 
\left. -
\frac{1-2V(a)-aV'(a)-\frac{b_-(a)+ab'_-(a)}{2a}}{\sqrt{1-\frac{b_{-}(a)}{a}-2V(a)}}
-
\sqrt{1-\frac{b_{-}(a)}{a}-2V(a)} \; a\Phi'_{-}(a)
\right],
\nonumber\\
\fl
&&
\label{gen-surfpressure2-onshell}
\end{eqnarray}
and
\begin{equation}
\fl
\Xi(a) 
=\frac{1}{4\pi a}\, \left[\Phi'_+(a)\sqrt{1-\frac{b_+(a)}{a}-2V(a)} - \Phi'_-(a)\sqrt{1-\frac{b_-(a)}{a}-2V(a)}\right]\,.
\end{equation}
The three quantities $\{\sigma(a),\P(a),\Xi(a)\}$ (or equivalently $\{m_s(a),\P(a),\Xi(a)\}$) are related by the differential conservation law, so at most two of them are functionally independent.

\subsection{Linearized equation of motion}

Consider a linearization around an assumed static solution (at $a_0$) to the equation of motion 
${1\over2}\dot a^2 + V(a)=0$,  and so also a solution of $\ddot a = -V'(a)$.  Generally a Taylor
expansion of $V(a)$ around $a_0$ to second order yields
\begin{equation}
\fl
V(a)=V(a_0)+V'(a_0)(a-a_0)+\frac{1}{2}V''(a_0)(a-a_0)^2+O[(a-a_0)^3]
\,.   \label{linear-potential0}
\end{equation}
But since we are expanding around a static solution $\dot a_0=\ddot a_0 = 0$, we automatically have 
$V(a_0)=V'(a_0)=0$, so it is sufficient to consider
\begin{equation}
V(a)= \frac{1}{2}V''(a_0)(a-a_0)^2+O[(a-a_0)^3]
\,.   \label{linear-potential}
\end{equation}
The assumed static solution at $a_0$ is stable if and only if $V(a)$ has a local minimum at $a_0$, 
which requires $V''(a_{0})>0$. This will be our primary criterion for gravastar stability, though it 
will be useful to rephrase it in terms of more basic quantities.

For instance, it is extremely useful to express $m_s'(a)$ and $m_s''(a)$ by the following 
expressions:
\begin{equation}
\fl
m_s'(a) = + {m_s(a)\over a} + {a\over2} \left\{ 
{ (b_+(a)/a)'+2V'(a)\over\sqrt{1-b_+(a)/a-2V(a)}} 
-
{(b_-(a)/a)'+2V'(a)\over\sqrt{1-b_-(a)/a-2V(a)}} \right\},
\end{equation}
and
\begin{eqnarray}
\fl
m_s''(a) &=& \left\{ 
{ (b_+(a)/a)'+2V'(a)\over\sqrt{1-b_+(a)/a-2V(a)}} 
-
{(b_-(a)/a)'+2V'(a)\over\sqrt{1-b_-(a)/a-2V(a)}} \right\}
\nonumber\\
\fl
&&
+{a\over4} 
\left\{ 
{ [(b_+(a)/a)'+2V'(a)]^2\over[1-b_+(a)/a-2V(a)]^{3/2}} 
-
{[(b_-(a)/a)'+2V'(a)]^2\over[1-b_-(a)/a-2V(a)]^{3/2}} \right\}
\nonumber\\
\fl
&&
+{a\over2} 
\left\{ 
{ (b_+(a)/a)''+2V''(a)\over\sqrt{1-b_+(a)/a-2V(a)}} 
-
{(b_-(a)/a)''+2V''(a)\over\sqrt{1-b_-(a)/a-2V(a)}} \right\}.
\end{eqnarray}
Doing so allows us to easily study linearized stability, and to develop a simple inequality on 
$m_s''(a_0)$ by using the constraint $V''(a_0)>0$. Similar formulae hold for $\sigma'(a)$, $\sigma''(a)$, for $\P'(a)$, $\P''(a)$, and for $\Xi'(a)$, $\Xi''(a)$. In view of the redundancies coming from the relations $m_s(a) = 4\pi\sigma(a) a^2$ and the differential conservation law, the only interesting quantities are  $\Xi'(a)$, $\Xi''(a)$.

It is similarly useful to consider
\begin{equation}
\fl
4\pi \,\Xi(a)\, a = \left[\Phi'_+(a)\sqrt{1-\frac{b_+(a)}{a}-2V(a)} - \Phi'_-(a)\sqrt{1-\frac{b_-(a)}{a}-2V(a)}\right]\,.
\end{equation}
for which an easy computation yields:
{\small
\begin{eqnarray}
\fl
[4\pi\,\Xi(a)\,a]' &=& 
+ \left\{ 
\Phi_+''(a) \sqrt{1-b_+(a)/a-2V(a)} 
- 
\Phi_-''(a) \sqrt{1-b_-(a)/a-2V(a)} \right\}
\nonumber
\\
\fl
&&
- {1\over2} \left\{ 
\Phi_+'(a) { (b_+(a)/a)'+2V'(a)\over\sqrt{1-b_+(a)/a-2V(a)}} 
- 
\Phi_-'(a){(b_-(a)/a)'+2V'(a)\over\sqrt{1-b_-(a)/a-2V(a)}} \right\},
\nonumber\\
\fl
&&
\end{eqnarray}
}
and
{\small
\begin{eqnarray}
\fl
[4\pi\,\Xi(a)\,a]'' &=& \left\{ 
\Phi_+'''(a) \sqrt{1-b_+(a)/a-2V(a)} 
- 
\Phi_-'''(a) \sqrt{1-b_-(a)/a-2V(a)} \right\}
\nonumber\\
\fl
&&
- \left\{ 
\Phi_+''(a) { (b_+(a)/a)'+2V'(a)\over\sqrt{1-b_+(a)/a-2V(a)}} 
-
\Phi_-''(a){(b_-(a)/a)'+2V'(a)\over\sqrt{1-b_-(a)/a-2V(a)}} \right\}
\nonumber\\
\fl
&&
-{1\over4} 
\left\{ 
\Phi_+'(a) { [(b_+(a)/a)'+2V'(a)]^2\over[1-b_+(a)/a-2V(a)]^{3/2}} 
- 
\Phi_-'(a) {[(b_-(a)/a)'+2V'(a)]^2\over[1-b_-(a)/a-2V(a)]^{3/2}} \right\}
\nonumber\\
\fl
&&
-{1\over2} 
\left\{ 
\Phi_+'(a) { (b_+(a)/a)''+2V''(a)\over\sqrt{1-b_+(a)/a-2V(a)}} 
-
\Phi_-'(a) {(b_-(a)/a)''+2V''(a)\over\sqrt{1-b_-(a)/a-2V(a)}} \right\}.
\nonumber\\
\fl
&&
\end{eqnarray}
}
We shall now evaluate these quantities at the assumed stable solution $a_0$.

\subsection{The master equations}
In view of the above, to have a stable static solution at $a_0$ we must have:
\begin{equation}
m_s(a_0) = -a_0
\left\{ 
  \sqrt{ 1- {b_+(a_0)\over a_0} } -\sqrt{ 1- {b_-(a_0)\over a_0}}
\right\},
 \label{stable_ms}
\end{equation}
while
\begin{equation}
m_s'(a_0) = {m_s(a_0)\over 2 a_0} -{1\over2} \left\{ 
{ 1 - b_+'(a_0)\over\sqrt{1-b_+(a_0)/a_0}} 
-
{ 1-  b_-'(a_0)\over\sqrt{1-b_-(a_0)/a_0}} \right\}.
 \label{stable_dms}
\end{equation}
The \emph{inequality} one derives for $m_s''(a_0)$ is now trickier since the relevant expression contains two competing terms of opposite sign.  Provided $b_+(a_0)\geq b_-(a_0)$, which is equivalent to demanding $\sigma(a_0)\geq 0$, one derives
\begin{eqnarray}
m_s''(a_0) &\geq&
+{1\over4 a_0^3} 
\left\{ 
{ [b_+(a_0)- a_0 b_+'(a_0)]^2\over[1-b_+(a_0)/a_0]^{3/2}} 
- 
{ [b_-(a_0)- a_0 b_-'(a_0)]^2\over[1-b_-(a_0)/a_0]^{3/2}}
\right\}
\nonumber\\
&&
+{1\over2} 
\left\{ 
{b_+''(a_0)\over\sqrt{1-b_+(a_0)/a_0}} 
-
{b_-''(a_0)\over\sqrt{1-b_-(a_0)/a_0}} \right\}. 
  \label{stable_ddms1}
\end{eqnarray}
However if $b_+(a_0)\leq b_-(a_0)$ the direction of the inequality is reversed.
This last formula in particular translates the stability condition $V''(a_0)\geq0$ into a rather 
explicit and not too complicated inequality on $m_s''(a_0)$, one that can in particular cases be 
explicitly checked with a minimum of effort.

In the absence of external forces this inequality is the only stability constraint one requires.  However, once one has external forces ($\Xi\neq 0$ which requires $\Phi_\pm\neq 0$),  there is additional information:
\begin{eqnarray}
\fl
\left.[4\pi\,\Xi(a)\,a]'\right|_{a_0} &=& 
+ \left.\left\{ 
\Phi_+''(a) \sqrt{1-b_+(a)/a} - 
\Phi_-''(a) \sqrt{1-b_-(a)/a} \right\}\right|_{a_0}
\nonumber
\\
\fl
&&
- {1\over2} \left.\left\{ 
\Phi_+'(a) { (b_+(a)/a)'\over\sqrt{1-b_+(a)/a}} - 
\Phi_-'(a){(b_-(a)/a)'\over\sqrt{1-b_-(a)/a}} \right\}\right|_{a_0}.
\end{eqnarray}
Provided $\Phi'_+(a_0)/\sqrt{1-b_+(a_0)/a_0} \geq \Phi'_-(a_0)/\sqrt{1-b_-(a_0)/a_0} $, we have
\begin{eqnarray}
\fl
\left.[4\pi\,\Xi(a)\,a]''\right|_{a_0} &\leq& \left.\left\{ 
\Phi_+'''(a) \sqrt{1-b_+(a)/a} - 
\Phi_-'''(a) \sqrt{1-b_-(a)/a} \right\}\right|_{a_0}
\nonumber\\
\fl
&&
- \left.\left\{ 
\Phi_+''(a) { (b_+(a)/a)'\over\sqrt{1-b_+(a)/a}} - 
\Phi_-''(a){(b_-(a)/a)'\over\sqrt{1-b_-(a)/a}} \right\}\right|_{a_0}
\nonumber\\
\fl
&&
-{1\over4} 
\left.\left\{ 
\Phi_+'(a) { [(b_+(a)/a)']^2\over[1-b_+(a)/a]^{3/2}} -
\Phi_-'(a) {[(b_-(a)/a)']^2\over[1-b_-(a)/a]^{3/2}} \right\}\right|_{a_0}
\nonumber\\
\fl
&&
-{1\over2} 
\left.\left\{ 
\Phi_+'(a) { (b_+(a)/a)''\over\sqrt{1-b_+(a)/a}} -
\Phi_-'(a) {(b_-(a)/a)''\over\sqrt{1-b_-(a)/a}} \right\}\right|_{a_0}.
   \label{stability_Xi}
\end{eqnarray}
If $\Phi'_+(a_0)/\sqrt{1-b_+(a_0)/a_0} \leq \Phi'_-(a_0)/\sqrt{1-b_-(a_0)/a_0} $ then the direction of the inequality is reversed.
Note that these last two equations are entirely vacuous in the absence of external forces, which is why they have not appeared in the literature until now.

\section{Specific gravastar models}\label{secIII}

In discussing specific gravastar models one now ``merely'' needs to apply the general 
formalism described above. Up to this stage we have kept the formalism as general as possible 
with a view to future applications, but we shall now focus on some more specific situations.

\subsection{Schwarzschild exterior, de~Sitter interior}\label{specific_example1}

The traditional gravastar, first considered in \cite{gr-qc/0012094} and \cite{gr-qc/0109035}, has a Schwarzschild exterior (with $b_+(r)=2M$ and $\Phi_+(r)=0$) 
and a de Sitter interior (with $b_-(r) = r^3/R^2$ and $\Phi_-(r)=0$), but with a complicated transition layer. Thin-shell Schwarzschild-de~Sitter gravastars were first explicitly discussed in~\cite{gr-qc/0310107}. The parameters are 
chosen such that the transition layer is located at some $2M < a < R$. (So the transition 
layer is situated outside the region where the Schwarzschild event horizon would normally  
form, and inside the region where the de Sitter cosmological horizon would form). One 
normally is rather noncommittal regarding the physics of the transition region, however, for the present case, the surface stresses of the thin shell are given by 
\begin{eqnarray}\label{sigmaP}
\sigma&=&-\frac{1}{4\pi a}\left[
 \sqrt{1-\frac{2M}{a}+\dot{a}^{2}}
-\sqrt{1-\frac{a^2}{R^2}+\dot{a}^{2}}
\right],
   \label{sigmaSdS1}
\\
\P&=&\frac{1}{8\pi a}\left[
\frac{1+\dot{a}^2+a\ddot{a}-\frac{M}{a}}{\sqrt{1-\frac{2M}{a}+\dot{a}^{2}}}     
 -
\frac{1+\dot{a}^2+a\ddot{a}-2\frac{a^2}{R^2}}{\sqrt{1-\frac{a^2}{R^2}+\dot{a}^{2}}}
\right].
\end{eqnarray}
The external forces vanish ($\Xi=0$), as $\Phi_\pm =0$, and $\sigma>0$ for $a(\tau)< (2M R^2)^{1/3}$, as can be readily verified from equation (\ref{sigmaSdS1}).
To have a stable static solution at $a_0$ we must have:
\begin{eqnarray}
\sigma(a_0)&=&-\frac{1}{4\pi a_0}\left[
 \sqrt{1-\frac{2M}{a_0}}
-\sqrt{1-\frac{a_0^2}{R^2}}
\right], \label{sigmaSchw}
\\
\P(a_0)&=&\frac{1}{8\pi a_0}\left[
\frac{1-\frac{M}{a_0}}{\sqrt{1-\frac{2M}{a_0}}}     
 -
\frac{1-2\frac{a_0^2}{R^2}}{\sqrt{1-\frac{a_0^2}{R^2}}}
\right], \label{PSchw}
\end{eqnarray}
with $2M<a_0<R$, a situation which has already been extensively analysed in Section \ref{secII_srtress_estimates}. As pointed out in the general discussion, $\sigma$ takes finite values with different 
signs in the endpoints of the range between $2M$ and $R$, being positive for $2M<a_0< (2M R^2)^{1/3}$ and 
negative for $(2M R^2)^{1/3} < a_0 < R$. The surface pressure $\P$ tends to 
$+\infty$ when $a_0$ approaches $2M$ ($R$) from the right (left). It can be seen that $\P$ never 
vanishes in this interval, and that its derivative with respect to $a_0$ tends to $-\infty$ 
($+\infty$) when $a_0$ goes to $2M$ ($R$) from the right (left). That is, 
$\P\left(a_0\right)$ evolves from infinitely large values, to a minimum non-vanishing value, 
before then again going to infinity.

\begin{figure}[!htb]
  \centering
  \includegraphics[width=3 in]{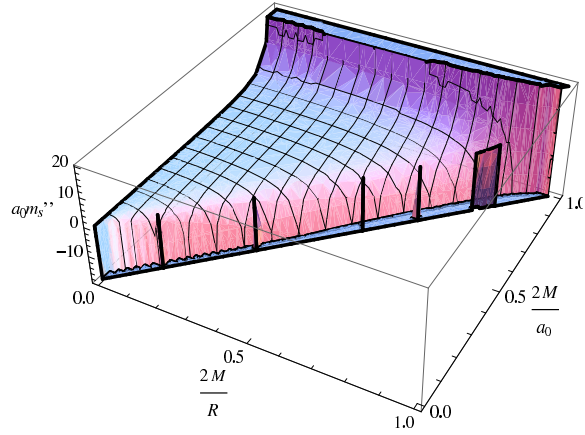}
  \caption{Schwarzschild-de~Sitter gravastar: The function $a_0\,m_s''$ in the case that $V=0$. 
  The surface is only defined for values of $a_0<R$ and $2M<a_0$, as 
  expected. Models producing a function $a_0\,m_s''$ above this surface would be in the 
  stability region.}
  \label{SchdS}
\end{figure}

Now consider the mass
\begin{equation}
m_s(a_0) = -a_0
\left\{ 
  \sqrt{ 1- {2M\over a_0} } -\sqrt{ 1- {a_0^2\over R^2}}
\right\},
\end{equation}
while
\begin{equation}
m_s'(a_0) = -\left\{ \frac{1-M/a_0}{\sqrt{1-2M/a_0}}-\frac{1-2a_0^2/R^2}{\sqrt{1-a_0^2/R^2}} \right\}.
\end{equation}
The inequality one derives for $m_s''(a_0)$ is now trickier since the relevant 
expression contains two competing terms of opposite sign. However, considering
that the physical solution should have $\sigma>0$ (equivalent to $a_0< (2M R^2)^{1/3}$), 
one finds that stability requires
\begin{eqnarray}
\fl
m_s''(a_0) &\geq&
+{1\over4 a_0^3} 
\left\{ 
{ [2M]^2\over[1-2M/a_0]^{3/2}} 
- 
{ [2a_0^3/R^2]^2\over[1-a_0^2/R^2]^{3/2}}
\right\}
-
\left\{ 
{3a_0/R^2\over\sqrt{1-a_0^2/R^2}} \right\}.
\end{eqnarray}
We can recast this as 
\begin{eqnarray}
a_0\, m_s''(a_0) &\geq&
{ \left(M/a_0\right)^2\over[1-2M/a_0]^{3/2}} 
- 
{ \left(a_0/R\right)^2 \; \left[3-2\left(a_0/R\right)^2\right] \over[1-\left(a_0/R\right)^2]^{3/2}}. 
\end{eqnarray}
In figure~\ref{SchdS} we show the surface which is produced when this inequality saturates,
where the stability region are represented above this surface. It is interesting to note that 
the possible positions of a static thin shell were studied in reference~\cite{gr-qc/0310107}. As 
those solutions, $a_0$, were obtained by considering a particular equation of state for the 
matter on the shell, and $\sigma$ and $\P$ are independent of $V''$, then the solutions would 
be the same for the zero potential as for the linearized potential. The only difference 
between the two models is that whereas if $V=0$ then the solution would be on the surface 
depicted in figure~\ref{SchdS}, if we consider the linearized potential then the solutions
should be in the region above the surface in order to assure stability of the static solution 
(which is equivalent to demanding $V''\left(a_0\right)>0$).

\subsection{Bounded excursion gravastars}

\begin{figure}[!htb]
  \centering
  \includegraphics[angle=-90,width=4.5 in]{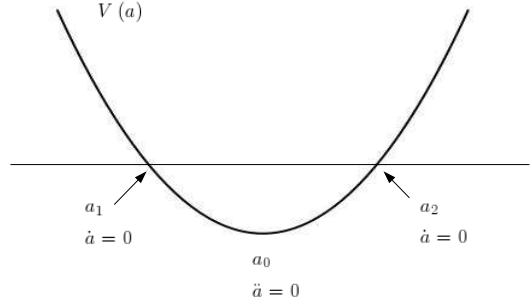}
  \caption{Bounded excursion potential obtained by deforming the linearized potential. $V(a)$ 
  vanishes at $a_{1,2}$, with $a_1<a_0<a_2$, being $a_0$ the minimum of the potential. The 
  shell expands from $a_1$ to $a_2$, and then it contracts again due to the nonvanishing 
  acceleration at this point.}
  \label{V}
\end{figure}

On the other hand, once one obtains a static solution at $a_0$ by requiring a specific 
equation of state (for example, stiff matter on the shell $\sigma=\P$), and verifies the 
stability by checking that the second derivative of the mass of the thin shell is in the 
stability region, it is easy to obtain dynamic ``stable'' solutions of the `bounded excursion' 
type by deforming the linearized potential. Thus, considering $V(a)=\frac{\gamma^2}{2}\left(a-
a_0\right)^2-{\epsilon^2\over2}$, with $\epsilon$ sufficiently small, one can obtain the 
equation of motion of the shell; this is
\begin{equation}
a(\tau)=a_0+\frac{\epsilon}{\gamma}\sin\left[\gamma\left(\tau-\tau_0\right)\right].
\end{equation}
Therefore, the shell expands from a minimum size with $a_1=a_0-\epsilon/\gamma$ to a maximum 
size corresponding to $a_2=a_0+\epsilon/\gamma$. It then contracts to $a_1$, starting a new 
cycle of evolution after reaching this value. Now, it can be clearly understood what we mean 
with $\epsilon$ sufficiently small, because this behavior makes sense for a stable gravastar
only if $a_1>2 M$ and $a_2<R$, which implies $\epsilon<\gamma \left(a_0-2 M\right)$ and 
$\epsilon<\gamma \left(R-a_0\right)$, respectively. In figure~\ref{V} we show the behavior of 
the potential, which is related with $\dot{a}$ through the equation of motion, as a function 
of $a$. This potential vanishes at $a_1$ and $a_2$, where $\dot{a}=0$ but $\ddot{a}=\pm 
\epsilon\gamma$, respectively. This acceleration imposes that the shell rolls down the 
potential when it has reached both its minimum and maximum sizes. The shell continues evolving 
when its radius takes the value $a_0$ because at this point $\dot{a}\neq 0$. 

Note that since
\begin{equation}
\frac{{\rm d}\tau}{{\rm d} t_\pm}=\frac{1-b_\pm(a)/a}{\sqrt{1-b_\pm(a)/a+\dot{a}^2}},
\end{equation}
one would have, in general, a different equation of motion of the shell in terms of the time 
coordinate of the exterior geometry and the time coordinate of the interior geometry, that is 
$a(t_+)\neq a(t_-)$. Nevertheless, as $\tau$ can be considered as a parameter in both regions 
of the space, the shell radius would always be bounded by $a_1$ and $a_2$. Thus, if $a_1$ and 
$a_2$ can be reached at a finite $t_\pm$, then the shell would be vibrating, although with a 
different kind of vibration as seen using $t_+$ or $t_-$. (See, for instance,~\cite{Gil}.) 

\begin{figure}[!htb]
  \includegraphics[width=2.9 in]{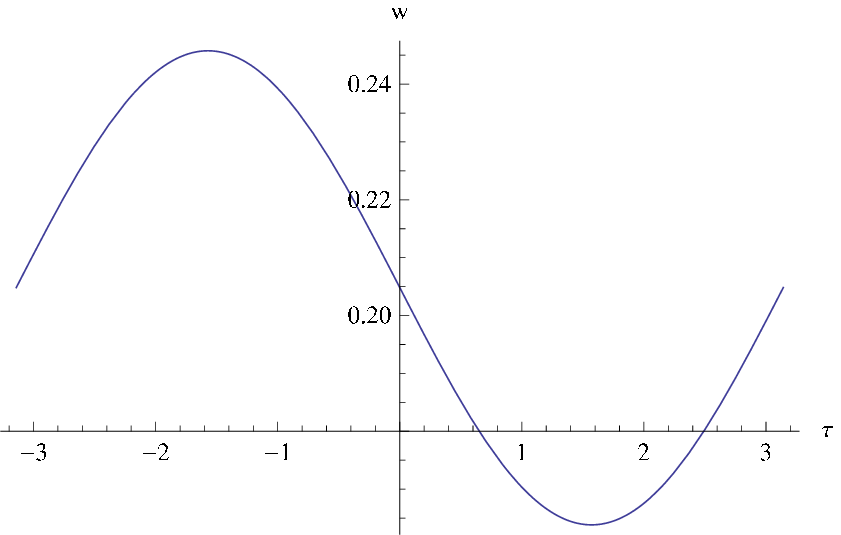}
  \includegraphics[width=2.9 in]{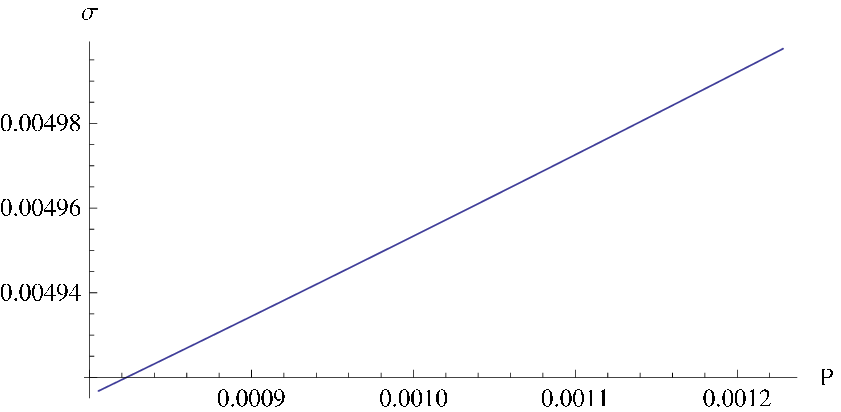}
  \caption{Bounded excursion: We consider the deformation $\epsilon=0.01$ of a stable static solution with dust 
  matter on the shell, $M=1$, $a_0=4.178821374980832$, $R=22.3607$ and $\gamma=1$. In the left 
  figure we show the evolution of $w$ in terms of $\tau$. It decreases from its maximum value 
  to its minimum value from $\tau_1=\tau(a=a_1)$ to $\tau_2=\tau(a=a_2)$ and then increases to
  its maximum value for the next cycle. Evolution of the energy density in terms of the 
  pressure is depicted in the right figure. Consideration of more than one cycle would 
  lead to the same graphic.}
  \label{w}
\end{figure}

Finally, we should comment on some features of the material on the shell. Although we are 
deforming a solution corresponding to a particular kind of material on the shell, that is with 
a given equation of state parameter relating $\sigma$ and $\P$, the corresponding dynamic
solution would not have a constant equation of state parameter, at least in the general case,
because the surface stresses, $\sigma(a)$ and $\P(a)$, have a different dependence on the 
trajectory (see 
equations~(\ref{sigmaP})). Therefore, the shell would be filled by material changing its
behavior during the evolution of the shell. Moreover, even if the original static solution 
corresponds to a material on the shell fulfilling the energy conditions, its dynamic 
generalization could easily violate those conditions at some stage of the evolution of the 
shell. Thus, one should carefully study that this is not the case in each particular model. In 
figure~\ref{w} we have depicted the equation of state parameter ($w=\P/\sigma$) as a function 
of time, and the energy density on the shell as a function of the pressure for a particular 
model obtained by deforming a stable static solution with dust matter on the shell. It can be 
seen that the pressure on the dynamic shell never vanishes, although the solution is obtained
by deforming one with $\P(a_0)=0$.

\subsection{Close to critical: Non-extremal}

Another common feature of gravastars is that the transition layer is typically taken to be 
``close'' to horizon formation. That is $b_\pm(a)/a \approx 1$. In fact, it is useful to take 
a linear approximation
\begin{equation}\label{expansion}
b_\pm(a)/a = 1 \mp \gamma_\pm(a-R_\mp) + \mathcal{O}([a-R_\pm]^2)\,,
\end{equation}
where $R_-$ is where the ``black hole'' horizon would have formed in the exterior spacetime, 
$R_+$ is where the ``cosmological'' horizon would have formed in the interior spacetime.
We take the transition layer to be at a position $a$ such that $R_-< a < R_+$, and $\gamma_\pm>0$
because we are (for now) avoiding ``extremal'' geometries. Dismissing second order terms and considering vanishing external forces
($\Phi_\pm(r)=0$), we have
\begin{eqnarray}
\fl
\sigma&\simeq&-\frac{1}{4\pi a}\left[
 \sqrt{\gamma_+\left(a-R_-\right)+\dot{a}^{2}}
-\sqrt{\gamma_-(R_+-a)+\dot{a}^{2}}
\right],
\\
\fl
\P&\simeq&\frac{1}{8\pi a}\left[
\frac{\dot{a}^2+a\ddot{a}+\gamma_+\left(3 a/2-R_-\right)}{\sqrt{\gamma_+\left(a-R_-\right)+\dot{a}^{2}}}     
 -
\frac{\dot{a}^2+a\ddot{a}+\gamma_-\left(R_+-3 a/2\right)}{\sqrt{\gamma_-\left(R_+-a\right)+\dot{a}^{2}}}
\right].
\end{eqnarray}
Therefore, $\sigma(a)>0$ for $a(\tau)<\left(\gamma_+R_-+\gamma_-R_+\right)/(\gamma_++\gamma_-)$.

This ``close to the horizon'' approximation would be valid only if the trajectory of the 
transition layer, which can be obtained from the equation of motion (or equivalently 
considering some equation of state), is always in the region $R_-\lesssim a\lesssim R_+$.
In order to analyze the accuracy of this approximation, we consider a stable static solution, 
implying
\begin{eqnarray}
\sigma\left(a_0\right)&\simeq&-\frac{1}{4\pi a_0}\left[
 \sqrt{\gamma_+\left(a_0-R_-\right)}
-\sqrt{\gamma_-(R_+-a_0)}
\right],
\\
\P\left(a_0\right)&\simeq&\frac{1}{8\pi a_0}\left[
\frac{\gamma_+\left(3 a_0/2-R_-\right)}{\sqrt{\gamma_+\left(a_0-R_-\right)}}     
 -
\frac{\gamma_-\left(R_+-3 a_0/2\right)}{\sqrt{\gamma_-\left(R_+-a_0\right)}}
\right].
\end{eqnarray}
Consider stiff matter on the shell, $\sigma\left(a_0\right)=\P\left(a_0\right)$. Therefore, we 
have
\begin{equation}\label{stiff}
\frac{\gamma_-^{1/2}\left(3R_+-7 a_0/2\right)}{\sqrt{R_+-a_0}}\simeq\frac{\gamma_+^{1/2}\left(7 a_0/2-3 R_-\right)}{\sqrt{a_0-R_-}}.
\end{equation}
Squaring both sides, and defining the dimensionless quantities $\alpha=a_0/R_-$, $\beta=R_-/R_+$, $\Gamma_+=\gamma_+R_-$,
and $\Gamma_-=\gamma_-R_-$, with $\alpha> 1$ and $0<\beta< 1$, one has
\begin{eqnarray}\label{equ}
&&\frac{49}{4}\left(\Gamma_++\Gamma_-\right)\alpha^3-\left[\frac{49}{4}\left(\Gamma_-+\frac{\Gamma_+}{\beta}\right)+21\left(\frac{\Gamma_-}{\beta}+\Gamma_+\right)\right]\alpha^2
\\ 
\nonumber
&&
\qquad+
\left[\frac{21}{\beta}\left(\Gamma_++\Gamma_-\right)+9\left(\frac{\Gamma_-}{\beta^2}+\Gamma_+\right)\right]\alpha
-\frac{9}{\beta}\left(\frac{\Gamma_-}{\beta}+\Gamma_+\right)\simeq0.
\end{eqnarray}
Thus, by considering the expansion of the background geometries close to where the horizon 
would be formed, we have reduced the problem of finding static and stable solutions (which 
usually involves some highly nontrivial equation) to solving a cubic, which can be done 
analytically.

Nevertheless, some comments are in order. In the first place, as the RHS of  
equation~(\ref{stiff}) is always positive, the solutions of equation~(\ref{equ}) would correspond to 
solutions of our problem only if $R_+>7 a_0/6$, that is $\beta<6/7$. In the second place,
one can consider that the approximation would not be accurate enough if $\left(a_0-
R_-\right)/R_-<1$ and $\left(R_+-a_0\right)/a_0<1$ are not satisfied; thus, the solutions 
would be reliable only if $\alpha<2$ and $1/4<\beta$. In summary, we should consider 
$1/4<\beta<6/7$ to solve equation~(\ref{equ}), and give physical meaning only to solutions with 
$1<\alpha<2$, if any. In fact, one can see that the results that can be obtained by using this 
approximation are compatible with those of a stiff matter gravastar with a Schwarzschild 
exterior and a de Sitter interior when $1<\alpha<2$ (see equation~(60) of reference~\cite{gr-qc/0310107} 
for the equation that must be solved in that case).

\begin{figure}[!htb]
  \includegraphics[width=2.9 in]{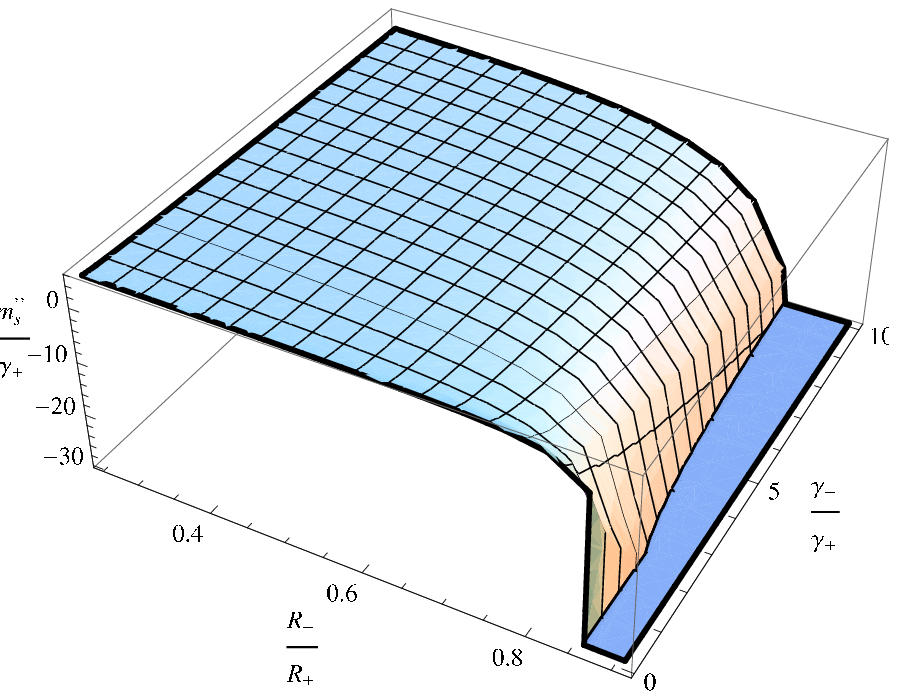}
  \includegraphics[width=2.9 in]{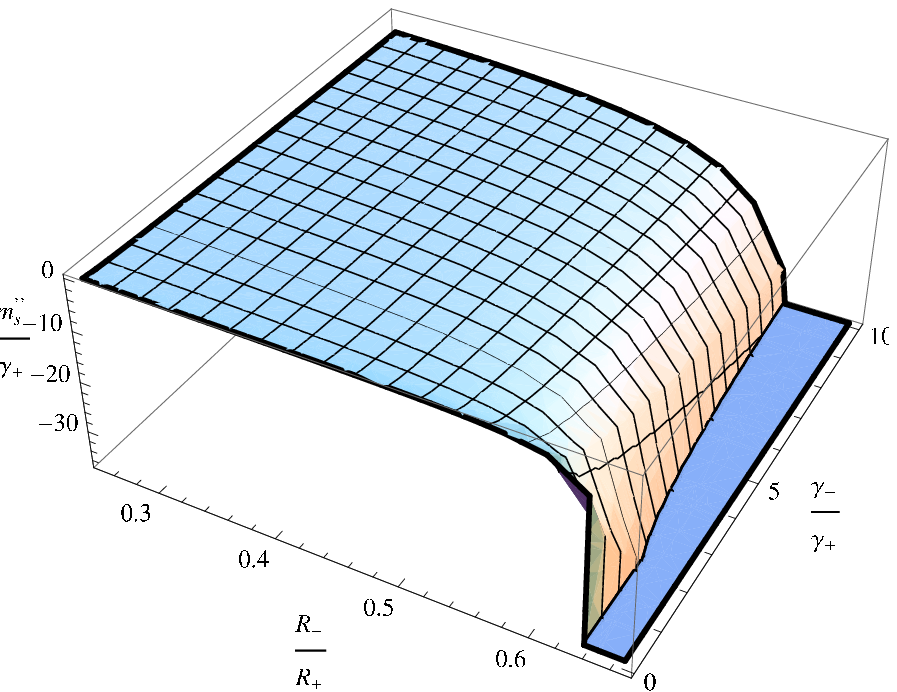}
  \caption{Close to critical (non-extremal): The left and right figures correspond to $a_0=1.1 R_-$ and $a_0=1.5 R_-$, 
  respectively. It can be noticed that the function is not defined for values of $R_+<a_0$, 
  since we must consider $R_-<a_0<R_+$. }
  \label{closehorizon}
\end{figure}

On the other hand, we can study the stability of the static solutions. In this case the 
inequality (\ref{stable_ddms1}), for $\sigma>0$, leads to
\begin{equation}
m_s''\left(a_0\right)\geq-\frac{\gamma_+^2\left(3a_0/4-R_-\right)}{\left[\gamma_+\left(a_0-
R_-\right)\right]^{3/2}}-\frac{\gamma_-^2\left(R_+-3a_0/4\right)}{\left[\gamma_-\left(R_+-
a_0\right)\right]^{3/2}}\,.
\end{equation}
It is more useful to consider $m_s''\left(a_0\right)/\gamma_+$, which can be written 
in terms of the dimensionless quantities previously introduced, and is given by
\begin{equation}
\frac{m_s''\left(a_0\right)}{\gamma_+}\geq-\frac{3\alpha/4-1}
{\left(\alpha-1\right)^{3/2}}-\sqrt{\frac{\Gamma_-}{\Gamma_+}}\frac{1/\beta-3\alpha/4}
{\left(1/\beta-\alpha\right)^{3/2}}.
\end{equation}
In figure~\ref{closehorizon} we have drawn this function for the case that the inequality 
saturates for two particular values of $a_0$. Thus the region above the surface corresponds to 
the stability region.

\subsection{Close to critical: Extremal}
We now consider a gravastar with a transition layer ``close'' to where the horizon 
would be expected to form, when both bulk geometries are ``extremal''. This implies that the coefficient of 
the first order term in the expansion (\ref{expansion}) vanishes. Then, assuming that the 
coefficient of the second order term is not vanishing, one has
\begin{equation}
b_\pm(a)/a = 1 - \gamma_\pm^ 2(a-R_\mp)^2 + \mathcal{O}([a-R_\pm]^3)\,,
\end{equation}
with $R_-< a < R_+$, and $\gamma_\pm>0$. Note particularly that we now have a different definition for $\gamma_\pm$.

Again considering the situation $\Phi_\pm(r)=0$, we can express the quantities related to the thin-shell as
\begin{eqnarray}
\fl
\sigma&\simeq&-\frac{1}{4\pi a}\left[
 \sqrt{\gamma_+^2\left(a-R_-\right)^2+\dot{a}^{2}}
-\sqrt{\gamma_-^2(R_+-a)^2+\dot{a}^{2}}
\right],
\\
\fl
\P&\simeq&\frac{1}{8\pi a}\left[
\frac{\dot{a}^2+a\ddot{a}+\gamma_+^2\left(2 a^2+R_-^2-3R_-a\right)}{\sqrt{\gamma_+^2\left(a-R_-\right)^2+\dot{a}^{2}}}-
\frac{\dot{a}^2+a\ddot{a}+\gamma_-^2\left(2 a^2+R_+^2-3R_+a\right)}{\sqrt{\gamma_-^2\left(R_+-a\right)^2+\dot{a}^{2}}}
\right].
\nonumber\\
\fl
&&
\end{eqnarray}
Thus, we recover that $a(\tau)<\left(\gamma_+R_-+\gamma_-R_+\right)/(\gamma_++\gamma_-)$ for $\sigma(a)>0$,
although $\gamma_\pm$ would be different than in the former case analyzed. Considering a static solution, one obtains
\begin{eqnarray}
\fl
\sigma\left(a_0\right)&\simeq&-\frac{1}{4\pi a_0}\left[
 \gamma_+\left(a_0-R_-\right)
-\gamma_-(R_+-a_0)
\right],
\\
\fl
\P\left(a_0\right)&\simeq&\frac{1}{8\pi a_0}\left[
\frac{\gamma_+\left(2 a_0^2+R_-^2-3R_-a_0\right)}{a_0-R_-}     
 -
\frac{\gamma_-\left(2 a_0^2+R_+^2-3R_+a_0\right)}{R_+-a_0}
\right],
\end{eqnarray}
where we have taken into account $R_-< a < R_+$, when simplifying. Following a similar 
procedure to that considered in the non-extremal case, one can see that a stiff matter shell 
must fulfil the equation
\begin{equation}\label{stiff2}
\frac{\gamma_-\left(4 a_0^2+3R_+^2-7R_+a_0\right)}{R_+-a_0}\simeq\frac{\gamma_+\left(4 a_0^2+3R_-^2-7R_-a_0\right)}{a_0-R_-},
\end{equation}
which, taking the same definition of $\Gamma_\pm$, $\beta$ and $\alpha$ introduced in the former case, leads to
\begin{eqnarray}\label{equ2}
&&4\left(\Gamma_++\Gamma_-\right)\alpha^3-\left[4\left(\Gamma_-+\frac{\Gamma_+}{\beta}\right)+7\left(\frac{\Gamma_-}{\beta}+\Gamma_+\right)\right]\alpha^2
\\ 
\nonumber
&&
\qquad+
\left[\frac{7}{\beta}\left(\Gamma_++\Gamma_-\right)+3\left(\frac{\Gamma_-}{\beta^2}+\Gamma_+\right)\right]\alpha
-\frac{3}{\beta}\left(\frac{\Gamma_-}{\beta}+\Gamma_+\right)\simeq0\,.
\end{eqnarray}
This is again a cubic equation.
However, whereas for non-extremal
geometries we have obtained equation~(\ref{equ}) by squaring equation~(\ref{stiff}), in this case that step was not necessary.
Thus, all solutions of equation~(\ref{equ2}) are also solutions of equation~(\ref{stiff2}).
Therefore, noticing the values of $\beta$ and $\alpha$ for which the approximation would be sufficiently accurate,
we should solve equation~(\ref{equ2}) for $1/4<\beta<1$ and then consider only the solutions with $1<\alpha<2$.

\begin{figure}[!htb]
  \centering
  \includegraphics[width=3 in]{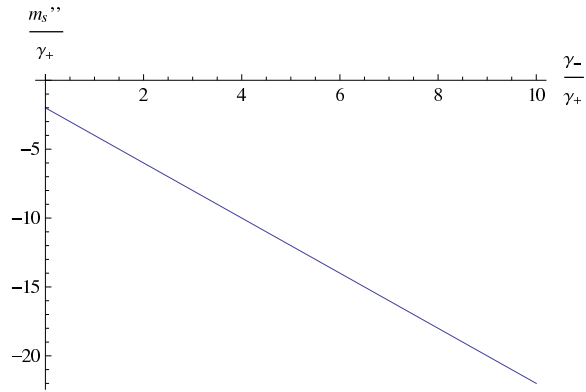}
  \caption{Close to critical (extremal): We show the function $m_s''\gamma_+$ in the case that $V=0$. The stability region would be above this curve, independent of the value of $a_0$. }
  \label{extremal}
\end{figure}

A special characteristic of these static solutions can be seen when studying their stability.
Considering $\sigma>0$, the inequality (\ref{stable_ddms1}) can be simplified to obtain
\begin{equation}
\frac{m_s''}{\gamma_+}\geq-2\left(1+\frac{\gamma_-}{\gamma_+}\right),
\end{equation}
which is independent of $a_0$. As the position $a_0$ can be obtained by considering a particular equation of state
for the material on the shell, this implies that the stability of the solution is independent of that equation of state\footnote[1]{This
interpretation could be reinforced by noticing that in this case $m_s''/\gamma_+$ depends only on the geometry and on the potential,
which, at the end of the day, implies a dependence on the geometry and on $\sigma$ (not on $\P$), if $\sigma\neq0$.}. In figure~\ref{extremal}, we show the behavior of this function.

Finally, some comments about the case with one non-extremal and one extremal background 
geometries are in order. In this case, for a stiff matter gravastar, one would obtain an equation with the RHS of equation~(\ref{stiff})
and the LHS of equation (\ref{stiff2}), or vice versa. Thus, after squaring,
the analogue of equation~(\ref{equ}) or (\ref{equ2}) would be a quintic. Therefore, the consideration of the close-to-horizon approximation in this situation would generally not significantly simplify the problem of obtaining static solutions, because we
are only reducing the polynomial equation by one degree (see also reference~\cite{gr-qc/0310107}).

\subsection{Charged dilatonic exterior, de Sitter interior}

An interesting solution to consider is that of an interior de Sitter spacetime, (where the 
metric functions are given by $b_-(r)=r^3/R^2$ and $\Phi_-(r)=0$), while the exterior spacetime 
is given by the dilaton black hole solution, which corresponds to an electric monopole. The latter de Sitter-charged dilatonic gravastar is considered here for the first time. In 
Schwarzschild coordinates, this exterior solution is described by~\cite{dirtyBH, dilaton1,dilaton2},
\begin{eqnarray}
\fl
ds^2_+ &=&- \left( 1 - {2M\over \beta +\sqrt{r^2+\beta^2} } \right)\, \d t^2
         + \left( 1 - {2M\over \beta +\sqrt{r^2+\beta^2} } \right)^{-1}
          {r^2\over r^2+\beta^2}\;\d r^2 
\nonumber  
\\   
\fl       
&& 
\vphantom{\Bigg|} \qquad\qquad + r^2(d\theta^2 + \sin^2\theta \; d\varphi^2) \,.
\end{eqnarray}
Here we have dropped the subscripts $+$ for notational convenience. The Lagrangian that describes this combined gravitational-electromagnetism-dilaton system is given by \cite{dirtyBH, dilaton1, dilaton2} 
\begin{equation}
{\cal L}=\sqrt{-g}\left\{-R/8\pi + 2 (\nabla\psi)^2 + e^{-2\psi} F^2/4\pi \right\}\,.
\end{equation}
(Note that the first charged dilatonic solutions, including
black hole solutions, were considered by Bronnikov \emph{et al.}~\cite{Bronn}).
The non-zero component of the electromagnetic tensor is given by $F_{\hat{t} \hat{r}} = Q/r^2$, and the dilaton field is given by $e^{2\psi} = 1 - {Q^2/ M(\beta+\sqrt{r^2+\beta^2})}$.
The parameter $\beta$ is defined by $\beta\equiv Q^2/2M$. 

In terms of the formalism developed in this paper, the metric functions of the exterior spacetime are
\begin{eqnarray}
b_+(r) &&=r \left[1-\left( 1 + {\beta^2\over r^2} \right)
                 \left( 1 - {2M\over \beta +\sqrt{r^2+\beta^2} } \right)\right], \\
\Phi_+(r) &&=-\frac{1}{2}\ln \left( 1 + {\beta^2\over r^2} \right).
\end{eqnarray}
An event horizon exists at $r_b = 2M\sqrt{1-\beta/M}$.
Note that $\Phi_+'(r)$ is given by
\begin{equation}
\Phi_+'(r)=\frac{\beta^2}{a(\beta^2+a^2)}\,,
\end{equation}
which is positive (and so satisfies the NEC) throughout the spacetime. Thus, taking into account that $\Phi_-=0$, we verify that the stability condition imposed by the presence of the flux term is governed by inequality (\ref{stability_Xi}).

\begin{figure}[!htb]
  \includegraphics[width=3.0 in]{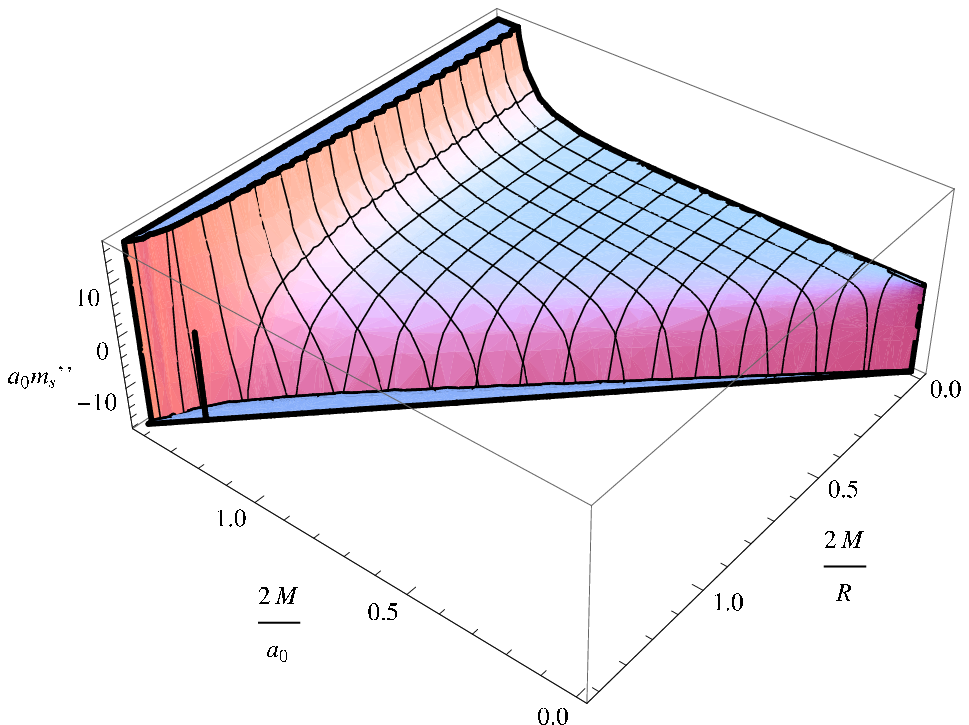}
  \includegraphics[width=3.0 in]{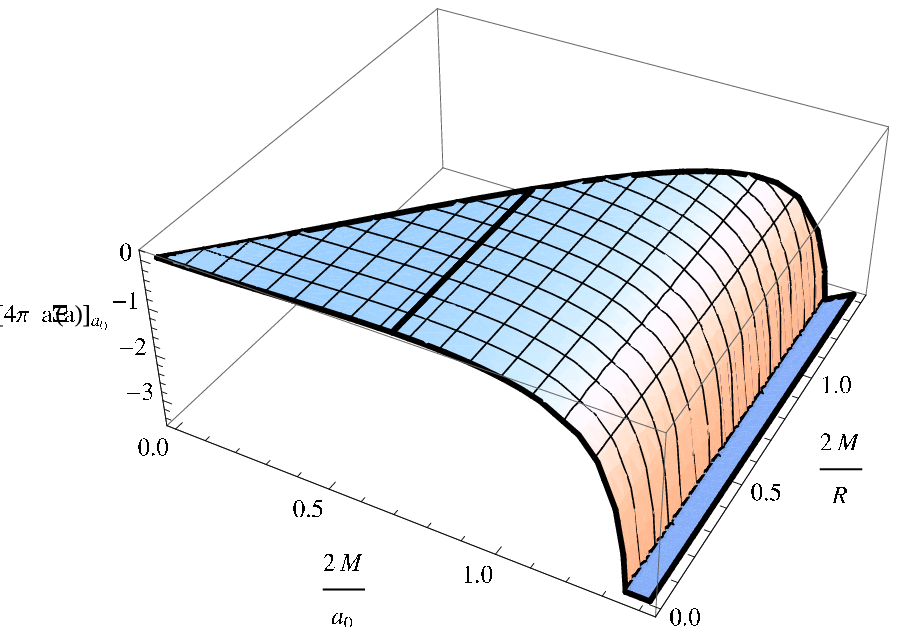}
  \caption{Dilaton--de~Sitter gravastar:  $\beta/M=0.5$. The left plot describes the 
  stability regions above the surface. The right plot describes the 
  stability regions below the surface. See the text for details.}
  \label{dilaton-blackhole1}
\end{figure}

\begin{figure}[!htb]
  \includegraphics[width=2.95 in]{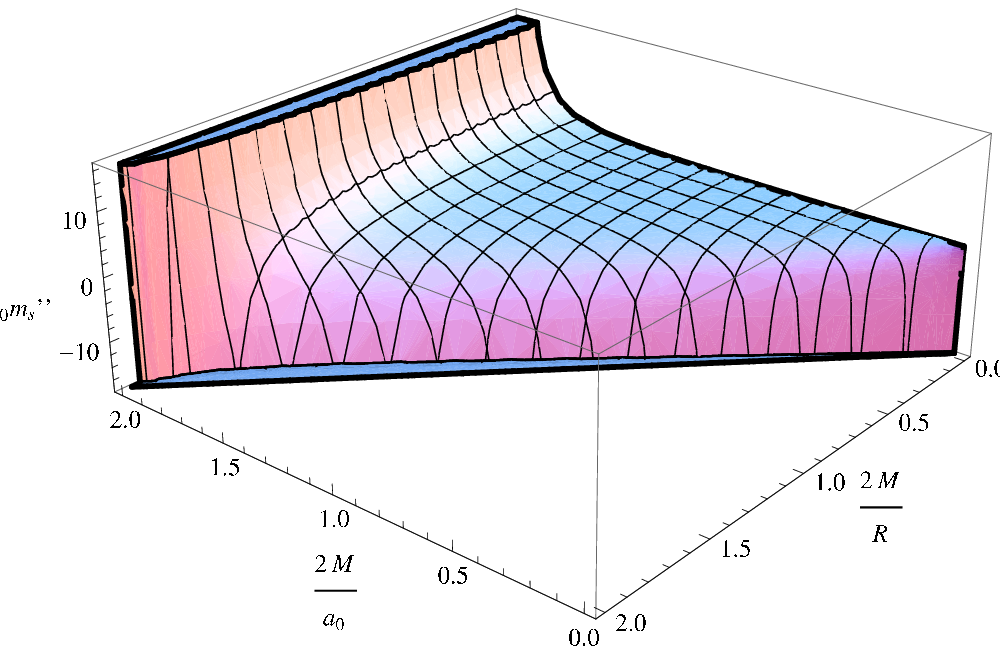}
  \includegraphics[width=2.95 in]{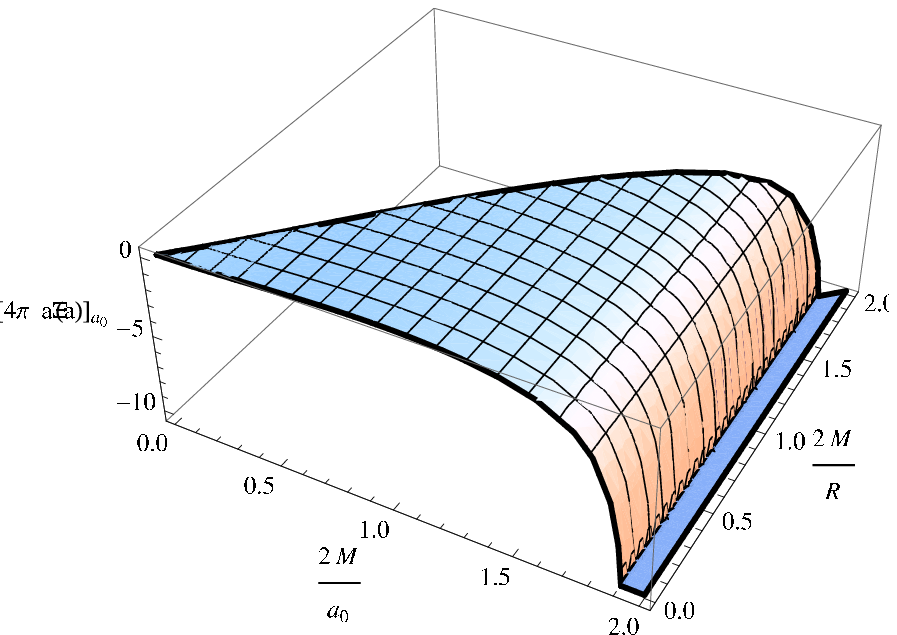}
  \caption{Dilaton--de~Sitter gravastar:  $\beta/M=0.75$. The left plot describes the 
  stability regions above the surface. The right plot describes the 
  stability regions below the surface. See the text for details.}
  \label{dilaton-blackhole2}
\end{figure}

\begin{figure}[!htb]
  \includegraphics[width=2.95 in]{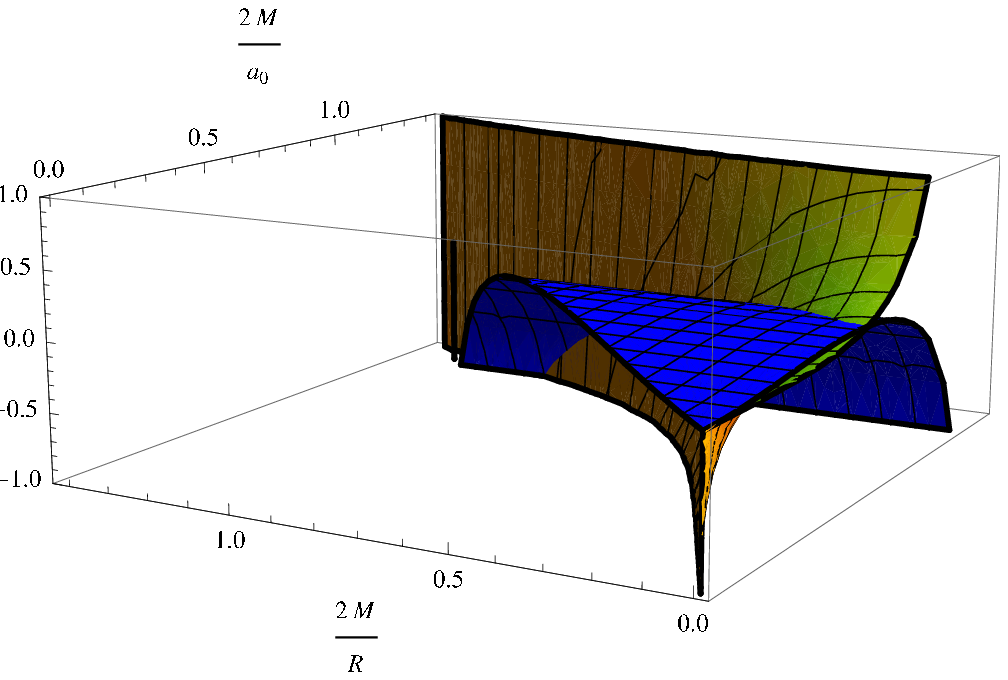}
  \includegraphics[width=2.95 in]{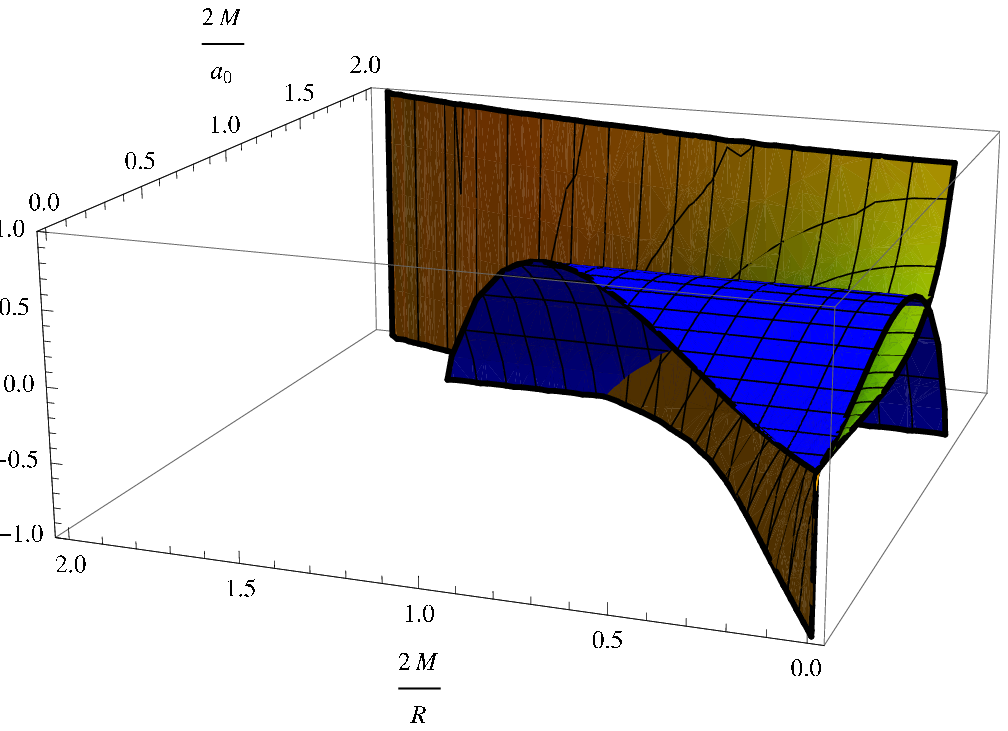}
  \caption{Dilaton--de~Sitter gravastar: The final stability regions for a thin shell dilaton gravastar for $\beta/M=0.5$ 
  and $\beta/M=0.75$ depicted in the left and right plots, respectively. The stability regions 
  are represented above the green surface, which is given by inequality (\ref{stable_ddms1}); 
  and below the blue surface, given by inequality (\ref{stability_Xi}). Thus the final 
  stability regions are given in between both surfaces. Note that the stability regions 
  increase for increasing values of $\beta/M$.}
  \label{dilaton-blackhole3}
\end{figure}

The thin shell is placed in the region $r_b < a < R$, so that $b_+(a)\geq 
b_-(a)$ (and consequently $\sigma>0$) is satisfied. Consequently the stability regions are governed by inequality 
(\ref{stable_ddms1}). The expressions for inequalities (\ref{stable_ddms1}) and 
(\ref{stability_Xi}) are extremely lengthy, so that rather than write them down explicitly, we 
will analyse the qualitative behaviour of the stability regions depicted in figures \ref{dilaton-blackhole1}--\ref{dilaton-blackhole3}. 

Consider as a first example the case for $\beta/M=1/2$, so that the stability 
regions, governed by the inequalities (\ref{stable_ddms1}) and (\ref{stability_Xi}), are 
depicted in figure \ref{dilaton-blackhole1}. The left plot describes the stability regions, 
governed by (\ref{stable_ddms1}), which lie above the surface. The right 
plot describes the stability regions depicted below the surface, provided by inequality (\ref{stability_Xi}). As a second case, consider the value $\beta/M=3/4$, depicted in figure \ref{dilaton-blackhole2}. We verify that the qualitative results are similar to the specific case of $\beta/M=1/2$, considered above. That is, the stability condition dictated by both the inequalities (\ref{stable_ddms1}) show that the stability regions decrease significantly as one approaches the black hole event horizon. The final stability regions are depicted in figure~\ref{dilaton-blackhole3}, and are shown in between the shaded regions. Note that the respective stability regions increase for increasing values of $\beta/M$.

\section{Summary and Discussion}\label{conclusion}

In this work, we have developed an extremely general and robust framework leading to the linearized 
stability analysis of dynamical spherically symmetric thin-shell gravastars, a framework that can quickly 
be adapted to wide classes of generic thin-shells. We have built on a companion paper, which 
analyzed thin-shell traversable wormholes~\cite{GLV}. We emphasize that mathematically there 
are a few strategic sign flips, which physically implies significant changes in the analysis. An 
important difference is the possibility of a vanishing surface energy density at certain 
specific shell radii, for the generic thin-shell gravastars considered in this work.
Due to the key sign flips, ultimately arising from the definition of the normals on the junction 
interface, the surface energy density is always negative for the thin-shell traversable 
wormholes considered in \cite{GLV}. We have also explored static gravastar configurations to 
some extent, and considered the generic qualitative behavior of the surface stresses. Relative 
to the conservation law of the surface stresses, we have analysed in great detail the most 
general case, widely ignored in the literature, namely, the presence of a flux term, 
corresponding to the net discontinuity in the (bulk) momentum flux which impinges on the 
shell. Physically, the latter flux term can be interpreted as the work done by external forces 
on the thin shell.

In the context of the linearized stability analysis we have reversed the logic flow typically considered 
in the literature and introduced a novel approach. More specifically, we have considered the 
surface mass as a function of the potential, so that specifying the latter tells us how much 
surface mass we need to put on the transition layer. This procedure implicitly makes demands 
on the equation of state of the matter residing on the transition layer and  
demonstrates in full generality that the stability of the gravastar is equivalent to choosing 
suitable properties for the material residing on the thin shell. We have applied the latter 
stability formalism to a number of specific cases, namely, to the traditional gravastar 
picture, where the transition layer separates an interior de Sitter space and an exterior 
Schwarschild geometry; bounded excursion; close-to-horizon models, in both the extremal 
and non-extremal regimes; and finally, to specific example of an interior de Sitter spacetime matched to 
a charged dilatonic exterior. This latter case is particularly interesting, as it 
involves the flux term.

In conclusion, by considering the matching of two generic static spherically symmetric 
spacetimes using the cut-and-paste procedure, we have analyzed the stability of thin-shell 
gravastars. The analysis provides a general and unified framework for simultaneously 
addressing a large number of gravastar models scattered throughout the literature.  As such we 
hope it will serve to bring some cohesion and focus to what is otherwise a rather disorganized 
and disparate collection of results. 

\ack

PMM acknowledges financial support from a FECYT postdoctoral mobility contract of the Spanish 
Ministry of Education through National Programme No.~2008--2011. NMG acknowledges financial 
support from CONACYT-Mexico. FSNL acknowledges financial support of the Funda\c{c}\~{a}o para 
a Ci\^{e}ncia e Tecnologia through Grants PTDC/\-FIS/\-102742/2008 and CERN/FP/116398/2010. MV  
acknowledges financial support from the Marsden Fund, administered by the Royal Society of New 
Zealand. 

\section*{References}



\begin{thebibliography}{99}



\bibitem{gr-qc/0012094} 
  G.~Chapline, E.~Hohlfeld, R.~B.~Laughlin and D.~I.~Santiago,
  ``Quantum phase transitions and the breakdown of classical general relativity'',  Int.\ J.\ 
  Mod.\ Phys.\ A\ {\bf 18}, 3587  (2003)  [gr-qc/0012094].  

\bibitem{gr-qc/0109035} 
  P.~O.~Mazur and E.~Mottola,
  ``Gravitational condensate stars: An alternative to black holes'',  gr-qc/0109035.  

  P.~O.~Mazur and E.~Mottola,
  ``Dark energy and condensate stars: Casimir energy in the large'',  gr-qc/0405111.  

  P.~O.~Mazur and E.~Mottola,
  ``Gravitational vacuum condensate stars'',  Proc.\ Nat.\ Acad.\ Sci.\ \ {\bf 101}, 9545  
  (2004)  [gr-qc/0407075].  

\bibitem{darkstar}
F.~S.~N.~Lobo,
  ``Stable dark energy stars'',
  Class.\ Quant.\ Grav.\  {\bf 23}, 1525-1541 (2006).
  [gr-qc/0508115].
  
 S.~S.~Yazadjiev,
  ``Exact dark energy star solutions'',  Phys.\ Rev.\ D\ {\bf 83}, 127501  (2011)  [arXiv:1104.1865 [gr-qc]].  
  
  R.~Chan, M.~F.~A.~da Silva and J.~F.~Villas da Rocha,
  ``On Anisotropic Dark Energy Stars'',  Mod.\ Phys.\ Lett.\ A\ {\bf 24}, 1137  (2009)  [arXiv:0803.2508 [gr-qc]].  

  A.~DeBenedictis, R.~Garattini and F.~S.~N.~Lobo,
  ``Phantom stars and topology change'',  Phys.\ Rev.\ D\ {\bf 78}, 104003  (2008)  [arXiv:0808.0839 [gr-qc]].  
  
 \bibitem{quasi-black-hole}
  J.~P.~S.~Lemos and O.~B.~Zaslavskii,
  ``Entropy of quasiblack holes'',
  Phys.\ Rev.\  D {\bf 81} (2010) 064012
  [arXiv:0904.1741 [gr-qc]];
  \\
 J.~P.~S.~Lemos and O.~B.~Zaslavskii,
  ``Black hole mimickers: regular versus singular behavior'',
  Phys.\ Rev.\  D {\bf 78} (2008) 024040
  [arXiv:0806.0845 [gr-qc]].

  
 
 \bibitem{monster}
  R.~D.~Sorkin, R.~M.~Wald and Z.~J.~Zhang,
 ``Entropy of selfgravitating radiation'',
  Gen.\ Rel.\ Grav.\  {\bf 13} (1981) 1127.
  \\
 S.~D.~H.~Hsu and D.~Reeb,
 ``Black hole entropy, curved space and monsters'',
  Phys.\ Lett.\  B {\bf 658} (2008) 244
  [arXiv:0706.3239 [hep-th]];
 \\
 S .~D.~H.~Hsu and D.~Reeb,
  ``Monsters, black holes and the statistical mechanics of gravity'',
  Mod.\ Phys.\ Lett.\  A {\bf 24} (2009) 1875
  [arXiv:0908.1265 [gr-qc]].
  
   \bibitem{black-star}
  C.~Barcel\'o, S.~Liberati, S.~Sonego and M.~Visser,
  ``Fate of gravitational collapse in semiclassical gravity'',
  Phys.\ Rev.\  D {\bf 77} (2008) 044032
  [arXiv:0712.1130 [gr-qc]];
  \\
 C.~Barcel\'o, S.~Liberati, S.~Sonego and M.~Visser,
 ``Revisiting the semiclassical gravity scenario for gravitational collapse'',
  AIP Conf.\ Proc.\  {\bf 1122} (2009) 99
  [arXiv: 0909.4157 [gr-qc]];
  \\
 C.~Barcel\'o, S.~Liberati, S.~Sonego and M.~Visser,
  ``Black stars, not holes'',
 Sci. Am. (Oct 2009) 8 pages;
  \\ 
  M.~Visser, C.~Barcel\'o, S.~Liberati and S.~Sonego,
  ``Small, dark, and heavy: But is it a black hole?'',
  arXiv: 0902.0346 [gr-qc].
  \\
  C.~Barcel\'o, S.~Liberati, S.~Sonego and M.~Visser,
  ``Hawking-like radiation from evolving black holes and compact horizonless objects,''
  JHEP {\bf 1102} (2011) 003
  [arXiv:1011.5911 [gr-qc]].
  \\
   C.~Barcel\'o, S.~Liberati, S.~Sonego and M.~Visser,
  ``Minimal conditions for the existence of a Hawking-like flux,''
  Phys.\ Rev.\ D {\bf 83} (2011) 041501
  [arXiv:1011.5593 [gr-qc]].
  \\
  M.~Visser,
  ``Black holes in general relativity,''
  arXiv:0901.4365 [gr-qc].
  
\bibitem{Dymnikova}
I. Dymnikova, ``Vacuum nonsingular black hole,'' Gen.
Rel. Grav. 24, 235 (1992);

I. Dymnikova, ``The algebraic structure of a cosmological
term in spherically symmetric solutions,'' Phys. Lett.
B472, 33-38 (2000) [arXiv:gr-qc/9912116];

I. Dymnikova, ``Cosmological term as a source
of mass,'' Class. Quant. Grav. 19 725-740 (2002)
[arXiv:gr-qc/0112052];

I. Dymnikova, ``Spherically symmetric space-time with
the regular de Sitter center,'' Int. J. Mod. Phys. D 12,
1015-1034 (2003) [arXiv:gr-qc/0304110];

I. Dymnikova and E. Galaktionov, ``Stability of a vacuum
nonsingular black hole,'' Class. Quant. Grav. 22 2331-
2358 (2005) [arXiv:gr-qc/0409049].
  
\bibitem{anisotropic} 
  C.~Cattoen, T.~Faber and M.~Visser,
  ``Gravastars must have anisotropic pressures'', 
   Class.\ Quant.\ Grav.\ \ {\bf 22}, 4189 (2005)  [gr-qc/0505137].  

\bibitem{anisotropic2} 
A.~DeBenedictis, D.~Horvat, S.~Ilijic, S.~Kloster and K.~S.~Viswanathan,
  ``Gravastar solutions with continuous pressures and equation of state'',  
  Class.\ Quant.\ Grav.\ \ {\bf 23}, 2303  (2006)  [gr-qc/0511097].  


\bibitem{gr-qc/0310107}
  M.~Visser, D.~L.~Wiltshire,
  ``Stable gravastars: An alternative to black holes?'',
  Class.\ Quant.\ Grav.\  {\bf 21 } (2004)  1135-1152.
  [gr-qc/0310107]. 
  
\bibitem{junction_formalism}
N. Sen, ``\"{U}ber die grenzbedingungen des schwerefeldes an
unsteig keitsfl\"{a}chen'', Ann. Phys. (Leipzig) {\bf 73}, 365
(1924);

K. Lanczos, ``Fl\"{a}chenhafte verteiliung der materie in der
Einsteinschen gravitationstheorie'', Ann. Phys. (Leipzig) {\bf
74}, 518 (1924);

G. Darmois, 
``Les equations de la gravitation einsteinienne'',
in \emph{M\'emorial des sciences math\'ematiques XXV}.
Fascicule XXV ch V (Gauthier-Villars, Paris, France, 1927);

S. O'Brien and J. L. Synge, 
``Jump conditions at discontinuity in general relativity.'',
Commun. Dublin Inst. Adv. Stud. Ser. A., no. 9 (1952) 1--20;

A. Lichnerowicz,
\emph{Th\'{e}ories Relativistes de la Gravitation et de l'Electromagnetisme},
 Masson, Paris (1955);

W. Israel, ``Singular hypersurfaces and thin shells in general
relativity'',   Nuovo Cimento {\bf 44}B, 1 (1966); and corrections
in {\it ibid.} {\bf 48}B, 463 (1966).


\bibitem{gravastar2} 
  N.~Bilic, G.~B.~Tupper and R.~D.~Viollier,
  ``Born-infeld phantom gravastars'',  JCAP\ {\bf 0602}, 013  (2006)  [astro-ph/0503427].  

  B.~M.~N.~Carter,
  ``Stable gravastars with generalised exteriors'',  Class.\ Quant.\ Grav.\ \ {\bf 22}, 4551  
  (2005)  [gr-qc/0509087].  

  F.~S.~N.~Lobo,
  ``Van der Waals quintessence stars'',  Phys.\ Rev.\ D\ {\bf 75}, 024023  (2007)  [gr-
  qc/0610118].  

  F.~S.~N.~Lobo and A.~V.~B.~Arellano,
  ``Gravastars supported by nonlinear electrodynamics'',  Class.\ Quant.\ Grav.\ \ {\bf 24}, 
  1069  (2007)  [gr-qc/0611083].  

  F.~S.~N.~Lobo,
  ``Stable dark energy stars: An alternative to black holes?'',  gr-qc/0612030.  

  D.~Horvat and S.~Ilijic,
  ``Gravastar energy conditions revisited'',  Class.\ Quant.\ Grav.\ \ {\bf 24}, 5637  (2007)  
  [arXiv:0707.1636 [gr-qc]].  

  R.~Chan, M.~F.~A.~da Silva and J.~F.~Villas da Rocha,
  ``Star Models with Dark Energy'',  Gen.\ Rel.\ Grav.\ \ {\bf 41}, 1835  (2009)  
  [arXiv:0803.3064 [gr-qc]].  

  P.~Rocha, A.~Y.~Miguelote, R.~Chan, M.~F.~da Silva, N.~O.~Santos and A.~Wang,
  ``Bounded excursion stable gravastars and black holes'',  JCAP\ {\bf 0806}, 025  (2008)  
  [arXiv:0803.4200 [gr-qc]].  

  D.~Horvat, S.~Ilijic and A.~Marunovic,
  ``Electrically charged gravastar configurations'',  Class.\ Quant.\ Grav.\ \ {\bf 26}, 
  025003  (2009)  [arXiv:0807.2051 [gr-qc]].  

  P.~Rocha, R.~Chan, M.~F.~A.~da Silva and A.~Wang,
  ``Stable and 'bounded excursion' gravastars, and black holes in Einstein's theory of 
  gravity'',  JCAP\ {\bf 0811}, 010  (2008)  [arXiv:0809.4879 [gr-qc]].  

  R.~Chan, M.~F.~A.~da Silva, P.~Rocha and A.~Wang,
  ``Stable Gravastars of Phantom Energy'',  JCAP\ {\bf 0903}, 010  (2009)  [arXiv:0812.4924 
  [gr-qc]].  

  B.~V.~Turimov, B.~J.~Ahmedov and A.~A.~Abdujabbarov,
  ``Electromagnetic Fields of Slowly Rotating Magnetized Gravastars'',  Mod.\ Phys.\ Lett.\ A\ 
  {\bf 24}, 733  (2009)  [arXiv:0902.0217 [gr-qc]].  

  R.~Chan, M.~F.~A.~da Silva and P.~Rocha,
  ``How the Cosmological Constant Affects the Gravastar Formation'',  JCAP\ {\bf 0912}, 017  
  (2009)  [arXiv:0910.2054 [gr-qc]].  

  F.~S.~N.~Lobo and R.~Garattini,
  ``Linearized stability analysis of gravastars in noncommutative geometry'',  arXiv:1004.2520 
  [gr-qc].  


  R.~Chan and M.~F.~A.~da Silva,
  ``How the Charge Can Affect the Formation of Gravastars'',  JCAP\ {\bf 1007}, 029  (2010)  
  [arXiv:1005.3703 [gr-qc]].  

  M.~E.~Gaspar and I.~Racz,
  ``Probing the stability of gravastars by dropping dust shells onto them'',  Class.\ Quant.\ 
  Grav.\ \ {\bf 27}, 185004  (2010)  [arXiv:1008.0554 [gr-qc]].  

  E.~Mottola,
  ``New Horizons in Gravity: The Trace Anomaly, Dark Energy and Condensate Stars'',  Acta 
  Phys.\ Polon.\ B\ {\bf 41}, 2031  (2010)  [arXiv:1008.5006 [gr-qc]].  

  R.~Chan, M.~F.~A.~da Silva and P.~Rocha,
  ``Gravastars and Black Holes of Anisotropic Dark Energy'',  Gen.\ Rel.\ Grav.\ \ {\bf 43}, 
  2223  (2011)  [arXiv:1009.4403 [gr-qc]].  

  A.~A.~Usmani, F.~Rahaman, S.~Ray, K.~K.~Nandi, P.~K.~F.~Kuhfittig, S.~.A.~Rakib and 
  Z.~Hasan,
  ``Charged gravastars admitting conformal motion'',  Phys.\ Lett.\ B\ {\bf 701}, 388  (2011)  
  [arXiv:1012.5605 [gr-qc]].  

  D.~Horvat, S.~Ilijic and A.~Marunovic,
  ``Radial stability analysis of the continuous pressure gravastar'',  Class.\ Quant.\ Grav.\ 
  \ {\bf 28}, 195008  (2011)  [arXiv:1104.3537 [gr-qc]].  

  R.~Chan, M.~F.~A.~da Silva, J.~F.~V.~da Rocha and A.~Wang,
  ``Radiating Gravastars'',  JCAP\ {\bf 1110}, 013  (2011)  [arXiv:1109.2062 [gr-qc]].  

\bibitem{thinshell2}
C. Barrab\`es and W. Israel, ``Thin shells in general relativity
and cosmology: The lightlike limit'', Phys. Rev. D {\bf 43}, 1129
(1991);

R. Mansouri and M. Khorrami, ``The equivalence of Darmois-Israel
and distributional-method for thin shells in general relativity'',
J. Math. Phys. {\bf 37}, 5672 (1996) [arXiv:gr-qc/9608029];

P. Musgrave and K. Lake, ``Junctions and thin shells in general
relativity using computer algebra I: The Darmois-Israel
formalism'', Class. Quant. Grav. {\bf 13} 1885 (1996)
[arXiv:gr-qc/9510052];

V. P. Frolov, M. A. Markov and V. F. Mukhanov, ``Black holes as
possible sources of closed and semiclosed worlds'', Phys. Rev. D
{\bf 41}, 383 (1990);

J. Fraundiener, C. Hoenselaers and W. Konrad, ``A shell around a
black hole'', Class. Quant. Grav. {\bf 7}, 585 (1990);

P. R. Brady, J. Louko and E. Poisson, ``Stability of a shell
around a black hole'', Phys. Rev. D {\bf 44}, 1891 (1991).



\bibitem{observations}
  A.~E.~Broderick and R.~Narayan,
  ``Where are all the gravastars? Limits upon the gravastar model from accreting black 
  holes'',  Class.\ Quant.\ Grav.\ \ {\bf 24}, 659  (2007)  [gr-qc/0701154 [GR-QC]].  

  C.~B.~M.~H.~Chirenti and L.~Rezzolla,
  ``How to tell a gravastar from a black hole'',  Class.\ Quant.\ Grav.\ \ {\bf 24}, 4191  
  (2007)  [arXiv:0706.1513 [gr-qc]].  

  J.~P.~S.~Lemos and O.~B.~Zaslavskii,
  ``Black hole mimickers: Regular versus singular behavior'',  Phys.\ Rev.\ D\ {\bf 78}, 
  024040  (2008)  [arXiv:0806.0845 [gr-qc]].  

  P.~Pani, V.~Cardoso, M.~Cadoni and M.~Cavaglia,
  ``Ergoregion instability of black hole mimickers'',  arXiv:0901.0850 [gr-qc].  

  S.~V.~Sushkov and O.~B.~Zaslavskii,
  ``Horizon closeness bounds for static black hole mimickers'',  Phys.\ Rev.\ D\ {\bf 79}, 
  067502  (2009)  [arXiv:0903.1510 [gr-qc]].  

  P.~Pani, E.~Berti, V.~Cardoso, Y.~Chen and R.~Norte,
  ``Gravitational wave signatures of the absence of an event horizon. I. Nonradial 
  oscillations of a thin-shell gravastar'',  Phys.\ Rev.\ D\ {\bf 80}, 124047  (2009)  
  [arXiv:0909.0287 [gr-qc]]. 

  T.~Harko, Z.~Kovacs and F.~S.~N.~Lobo,
  ``Can accretion disk properties distinguish gravastars from black holes?'',  Class.\ Quant.\ 
  Grav.\ \ {\bf 26}, 215006  (2009)  [arXiv:0905.1355 [gr-qc]].  


\bibitem{GLV}
N.~M.~Garcia, F.~S.~N.~Lobo and M.~Visser
  ``Generic spherically symmetric dynamic thin-shell traversable wormholes in standard 
  general relativity'',  arXiv:1112.2057 [gr-qc].  
  
\bibitem{voids}
C.~Barcel\'o and M.~Visser,
  ``Living on the edge: Cosmology on the boundary of Anti-de Sitter space,''
  Phys.\ Lett.\ B {\bf 482} (2000) 183
  [hep-th/0004056].
\\
C.~Barcel\;o and M.~Visser,
  ``Brane surgery: Energy conditions, traversable wormholes, and voids,''
  Nucl.\ Phys.\ B {\bf 584} (2000) 415
  [hep-th/0004022].
  
\bibitem{MTWbook}
C. W. Misner, K. S. Thorne and J. A. Wheeler, Gravitation (W. H.
Freeman and Company, New York, 1973).

\bibitem{EC-violations}
C.~Barcel\'o and M.~Visser,
  ``Twilight for the energy conditions?,''
  Int.\ J.\ Mod.\ Phys.\ D {\bf 11} (2002) 1553
  [gr-qc/0205066].
  \\
  M.~Visser,
  ``Gravitational vacuum polarization,''
  gr-qc/9710034.
  \\
  M.~Visser,
  ``Gravitational vacuum polarization. 1: Energy conditions in the Hartle-Hawking vacuum,''
  Phys.\ Rev.\ D {\bf 54} (1996) 5103
  [gr-qc/9604007].
  \\
  M.~Visser,
  ``Gravitational vacuum polarization. 2: Energy conditions in the Boulware vacuum,''
  Phys.\ Rev.\ D {\bf 54} (1996) 5116
  [gr-qc/9604008].
  \\
  M.~Visser,
  ``Gravitational vacuum polarization. 3: Energy conditions in the (1+1) Schwarzschild space-time,''
  Phys.\ Rev.\ D {\bf 54} (1996) 5123
  [gr-qc/9604009].
  \\
  M.~Visser,
  ``Gravitational vacuum polarization. 4: Energy conditions in the Unruh vacuum,''
  Phys.\ Rev.\ D {\bf 56} (1997) 936
  [gr-qc/9703001].


\bibitem{Visser}
M. Visser, \emph{Lorentzian Wormholes: From Einstein to Hawking},
(American Institute of Physics, New York, 1995).


\bibitem{thinshellWH}
M. Visser, ``Traversable wormholes: Some simple examples'', Phys.
Rev. D {\bf 39} 3182 (1989);

M. Visser, ``Traversable wormholes from surgically modified
Schwarzschild spacetimes'', Nucl. Phys. {\bf B 328} 203 (1989);

M. Visser, ``Quantum mechanical stabilization of Minkowski
signature wormholes'', Phys. Lett. B {\bf 242}, 24 (1990);

E.~Poisson and M.~Visser,
  ``Thin shell wormholes: Linearization stability'',
  Phys.\ Rev.\  D {\bf 52}, 7318 (1995)
  [arXiv:gr-qc/9506083];

E. F. Eiroa and G. E. Romero ``Linearized stability of charged
thin-shell wormoles'', Gen. Rel. Grav. {\bf 36} 651-659 (2004)
[arXiv:gr-qc/0303093];

F.~S.~N.~Lobo and P.~Crawford,
  ``Linearized stability analysis of thin shell wormholes with a cosmological
  constant'',
  Class.\ Quant.\ Grav.\  {\bf 21}, 391 (2004)
  [arXiv:gr-qc/0311002];
  
J.~P.~S.~Lemos and F.~S.~N.~Lobo,
  ``Plane symmetric thin-shell wormholes: Solutions and stability,''  Phys.\ Rev.\ D {\bf 78}, 
  044030 (2008)  [arXiv:0806.4459 [gr-qc]].  

  K.~A.~Bronnikov and A.~A.~Starobinsky,
  ``Once again on thin-shell wormholes in scalar-tensor gravity,''
  Mod.\ Phys.\ Lett.\  A {\bf 24}, 1559 (2009)
  [arXiv:0903.5173 [gr-qc]].





\bibitem{thinshellWH2}
J. P. S. Lemos, F. S. N. Lobo and S. Q. de Oliveira,
``Morris-Thorne wormholes with a cosmological constant'', Phys.
Rev. D {\bf 68}, 064004 (2003) [arXiv:gr-qc/0302049];

J.~P.~S.~Lemos and F.~S.~N.~Lobo, ``Plane symmetric traversable
wormholes in an anti-de Sitter background'', Phys.\ Rev.\ D {\bf
69}, 104007 (2004) [arXiv:gr-qc/0402099];

F. S. N. Lobo,
``Energy conditions, traversable wormholes and dust shells'',
Gen.\ Rel.\ Grav.\ \ {\bf 37} (2005) 2023
[arXiv:gr-qc/0410087];

F. S. N. Lobo, ``Phantom energy traversable wormholes'', Phys.
Rev. D {\bf 71}, 084011 (2005) [arXiv:gr-qc/0502099];

  F.~S.~N.~Lobo,
  ``Stability of phantom wormholes'',
  Phys.\ Rev.\  {\bf D71}, 124022 (2005).
  [gr-qc/0506001];

  F.~S.~N.~Lobo, P.~Crawford,
  ``Stability analysis of dynamic thin shells'',
  Class.\ Quant.\ Grav.\  {\bf 22}, 4869-4886 (2005).
  [gr-qc/0507063];
  
F.~S.~N.~Lobo,
  ``Exotic solutions in General Relativity: Traversable wormholes and 'warp drive' 
  spacetimes,''  arXiv:0710.4474 [gr-qc].  


\bibitem{gr-qc/0409018}
F. S. N. Lobo, ``Surface stresses on a thin shell surrounding a
traversable wormhole'', Class. Quant. Grav. {\bf 21} 4811 (2004)
[arXiv:gr-qc/0409018].
\bibitem{Abreu}
G.~Abreu and M.~Visser,
  ``Tolman mass, generalized surface gravity, and entropy bounds,''
  Phys.\ Rev.\ Lett.\  {\bf 105} (2010) 041302
  [arXiv:1005.1132 [gr-qc]].

\bibitem{Ishak}
M. Ishak and K. Lake, ``Stability of transparent spherically
symmetric thin shells and wormholes'', Phys. Rev. D {\bf 65}
044011 (2002) [arXiv:gr-qc/0108058].

\bibitem{Gil}
L.~C.~Barbado, C.~Barcel\'o, L.~J.~Garay and G.~Jannes,
  ``The Trans-Planckian problem as a guiding principle,''
  JHEP {\bf 1111} (2011) 112
  [arXiv:1109.3593 [gr-qc]].

\bibitem{dirtyBH}
  M.~Visser,
  ``Dirty black holes: Thermodynamics and horizon structure'',  
Phys.\ Rev.\ D\ {\bf 46}, 2445  (1992)  [arXiv:hep-th/9203057 [hep-th]].  
\bibitem{dilaton1}
G.~W.~Gibbons and K.~-i.~Maeda,
  ``Black Holes and Membranes in Higher Dimensional Theories with Dilaton Fields'',
  Nucl.\ Phys.\ B\ {\bf 298} (1988) 741.
\bibitem{dilaton2}
D.~Garfinkle, G.~T.~Horowitz and A.~Strominger,
``Charged black holes in string theory'',
  Phys.\ Rev.\ D\ {\bf 43} (1991) 3140
   [Erratum-ibid.\ D\ {\bf 45} (1992) 3888].

\bibitem{Bronn} 
  K.~A.~Bronnikov and G.~N.~Shikin,
  ``Interacting Fields in General Relativity,''  Izv.\ Vuz.\ Fiz.\  {\bf 1977N9}, 25 (1977).  
(English translation: Russ. Phys. J. v.20, No.9, p. 1138-1143, 1977).

\end{thebibliography}
\end{document}